%% file: a959final.tex
\documentclass{aa}  
\usepackage{graphicx}
\usepackage{txfonts}
\newcommand{\mincir}{\raise -2.truept\hbox{\rlap{\hbox{$\sim$}}\raise5.truept
\hbox{$<$}\ }}
\newcommand{\magcir}{\raise -2.truept\hbox{\rlap{\hbox{$\sim$}}\raise5.truept
\hbox{$>$}\ }}
\newcommand{\siml}{\raise -2.truept\hbox{\rlap{\hbox{$\sim$}}\raise5.truept
\hbox{$<$}\ }}
\newcommand{\simg}{\raise -2.truept\hbox{\rlap{\hbox{$\sim$}}\raise5.truept
\hbox{$>$}\ }}
\newcommand{\be}{\begin{equation}}
\newcommand{\ee}{\end{equation}}
\newcommand{\ba}{\begin{eqnarray}}
\newcommand{\ea}{\end{eqnarray}}
\newcommand {\kpc} {$\mathrm{h_{70}^{-1}}$ kpc $\;$}

\newcommand {\h} {$h_{70}^{-1}$ Mpc$\;$}
\newcommand {\hh} {$h_{70}^{-1}$ Mpc}
\newcommand {\hhh} {\;h_{70}^{-1} \mathrm{Mpc}}
\newcommand {\ks} {km~s$^{-1} \;$}
\newcommand {\kss} {km~s$^{-1}$}

\newcommand {\mqui} {$\times 10^{15}\;h_{70}^{-1}\;M_{\odot} \;$}
\newcommand {\mquii} {$\times 10^{15}\;h_{70}^{-1}\;M_{\odot}$}

\newcommand{\degree}{\ensuremath{\mathrm{^\circ}}}
\newcommand{\arcm}{\ensuremath{\mathrm{^\prime}\;}}

\newcommand{\arcmm}{\ensuremath{\mathrm{^\prime}}}

\newcommand{\dotarcs}{\,\rlap{\hbox{$\mathrm{^\prime\hskip-0.1em^\prime}$}}{\hbox{$.$}}\,}

\newcommand{\dotsec}{\,\rlap{\hbox{$\mathrm{^s}$}}{\hbox{$.$}}\,}

\begin{document}
   \title{Internal dynamics of the galaxy cluster Abell 959}

   \author{W. Boschin\inst{1,2}
          \and
R. Barrena\inst{3}
          \and
M. Girardi\inst{2,4}
}

   \offprints{W. Boschin, \email{boschin@tng.iac.es}}

   \institute{
	      Fundaci\'on Galileo Galilei - INAF, Rambla Jos\'e Ana 
              Fern\'andez Perez 7, E-38712 Bre\~na Baja (La Palma), 
              Canary Islands, Spain\\
\and 
Dipartimento di Astronomia of the Universit\`a degli
	      Studi di Trieste, via Tiepolo 11, I-34143 Trieste,
	      Italy\\ 
\and
Instituto de Astrof\'{\i}sica de Canarias,
	      C/V\'{\i}a L\'actea s/n, E-38205 La Laguna (Tenerife),
	      Canary Islands, Spain\\
\and 
INAF - Osservatorio Astronomico di Trieste, via Tiepolo 11, I-34143
	      Trieste, Italy\\ 
             }

\date{Received  / Accepted }

\abstract {The connection of cluster mergers with the presence of
  extended, diffuse radio sources in galaxy clusters is still being
  debated.}{We aim to obtain new insights into the internal dynamics
  of Abell 959, showing evidence of a diffuse radio source, analyzing
  velocities and positions of member galaxies.}{Our analysis is based
  on redshift data for 107 galaxies in the cluster field acquired at
  the Telescopio Nazionale Galileo. We also use photometric data from
  the Sloan Digital Sky Survey (Data Release 6). We combine galaxy
  velocities and positions to select 81 galaxies recognized as cluster
  members and determine global dynamical properties. We analyze the
  cluster searching for substructures by using the weighted gap
  analysis, the KMM method and the Dressler--Shectman statistics. We
  also study the 2D galaxy distribution in the field of the
  cluster. We compare our results with those from X--ray and
  gravitational lensing analyses.}{ We estimate a cluster redshift of
  $\left<z\right>=0.2883\pm0.0004$. We detect an NE high velocity group
  at 5\arcm from the cluster center with a relative line--of--sight
  (LOS) velocity of $\sim +1900$ \ks with respect to the main
  system. We also detect a central, dense structure elongated along
  the SE--NW direction likely connected with the two dominant galaxies
  and their surrounding cores. This elongated central structure is
  probably the trace of an old cluster merger. The LOS velocity
  dispersion of galaxies is very high. By excluding the NE clump we
  obtain $\sigma_{\rm V}=1025_{-75}^{+104}$ \ks and a virial mass
  $M(<R=1.48 \hhh)=1.15^{-0.19}_{+0.25}$ \mquii. Our results suggest
  that this cluster is forming along two main directions of mass
  accretion.}{Abell 959 is confirmed to show the typical
  characteristics of radio clusters; i.e., it is very massive and
  shows a young dynamical state. However, deeper radio observations
  are needed to clarify the nature of the diffuse radio emission in
  Abell 959 and its connection with the cluster internal dynamics.}

  \keywords{Galaxies: clusters: individual: Abell 959 -- Galaxies:
             clusters: general -- Galaxies: distances and redshifts }

   \maketitle
%
%________________________________________________________________

\section{Introduction}
\label{intr}

In the hierarchical scenario for large--scale structure formation,
clusters of galaxies are known to be unrelaxed structures. Instead, as
the cosmic time flows clusters are interested by perturbing merging
processes that constitute an essential ingredient of their
evolution. Much progress has been made in the past decade in the
observations of the signatures of merging processes (see Feretti et
al. \cite{fer02b} for a general review). A recent aspect of these
investigations is the possible connection of cluster mergers with the
presence of extended, diffuse radio sources: halos and relics. The
synchrotron radio emission of these sources demonstrates the existence
of large--scale cluster magnetic fields and of widespread relativistic
particles. It has been suggested that cluster mergers provide the
large amount of energy necessary for electron reacceleration up to
relativistic energies and for magnetic field amplification (Feretti
\cite{fer99}; Feretti \cite{fer02a}; Sarazin \cite{sar02}). However,
the precise radio halos/relics formation scenario is still being
debated since the diffuse radio sources are quite uncommon and only
recently can we study these phenomena on the basis of sufficient
statistics (few dozen clusters up to $z\sim 0.3$, e.g., Giovannini et
al. \cite{gio99}; see also Giovannini \& Feretti \cite{gio02}; Feretti
\cite{fer05}).

There is growing evidence of the connection between diffuse radio
emission and cluster merging based on X--ray data (e.g., B\"ohringer
\& Schuecker \cite{boh02}; Buote \cite{buo02}). Studies based on a
large number of clusters have found a significant relation between the
radio and the X--ray surface brightness (Govoni et al. \cite{gov01a},
\cite{gov01b}) and connections between the presence of
radio--halos/relics and irregular and bimodal X--ray surface
brightness distribution (Schuecker et al. \cite{sch01}).

Optical data are a powerful way to investigate the presence and the
dynamics of cluster mergers (e.g., Girardi \& Biviano \cite{gir02}),
too.  The spatial and kinematical analysis of member galaxies allow us
to detect and measure the amount of substructures and to identify and
analyze possible pre--merging clumps or merger remnants.  This optical
information is really complementary to X--ray information since
galaxies and intracluster medium react on different time scales
during a merger (see, e.g., numerical simulations by Roettiger et
al. \cite{roe97}). In this context we are conducting an intensive
observational and data analysis program to study the internal dynamics
of radio clusters by using member galaxies. Our program concerns both
massive clusters, where diffuse radio emissions are more frequently
found (e.g., Girardi et al. \cite{gir08} and refs. therein), and
low--mass galaxy systems (Boschin et al. \cite{bos08}\footnote{please
visit the web site of the DARC (Dynamical Analysis of Radio Clusters)
project: http://adlibitum.oat.ts.astro.it/girardi/darc.}).

During our observational program we conducted an intensive study of
the cluster \object{Abell 959} (hereafter A959). It is a fairly rich,
X--ray luminous, hot and massive Abell cluster: Abell richness class
$=1$ (Abell et al. \cite{abe89}); $L_\mathrm{X}$(0.1--2.4
keV)=13.25$\times 10^{44} \ h_{70}^{-2}$ erg\ s$^{-1}$ (Popesso et
al. \cite{pop04}, but correcting for our $<z>$, see below);
$T_\mathrm{X}= 6.95_{-1.33}^{+1.85}$ keV (ASCA data, Mushotzky \&
Scharf \cite{mus97}); a projected 2D aperture mass within a radius of
440\arcsec $M_{\rm ap}\sim1-2$ \mqui (Dahle et al. \cite{dah02}, in
their cosmology).

Optically, the cluster center is not dominated by any single galaxy --
it is classified as Bautz--Morgan class III (Abell et
al. \cite{abe89}) -- but it has a core region consisting of many
early--type galaxies of similar brightness (Dahle et
al. \cite{dah02}). The general SE--NW elongation of the A959 cluster
galaxy distribution is matched in X--rays (ROSAT/PSPC data; Dahle et
al. \cite{dah03}). Perpendicularly, the NE--SW direction is defined by
a clear NE external extension and a likely SW extension in the X--ray
emission (Dahle et al. \cite{dah03}) and by a possible very central
bimodal structure (Ota \& Mitsuda \cite{ota04}).

\begin{figure*}
\centering 
\includegraphics[width=18cm]{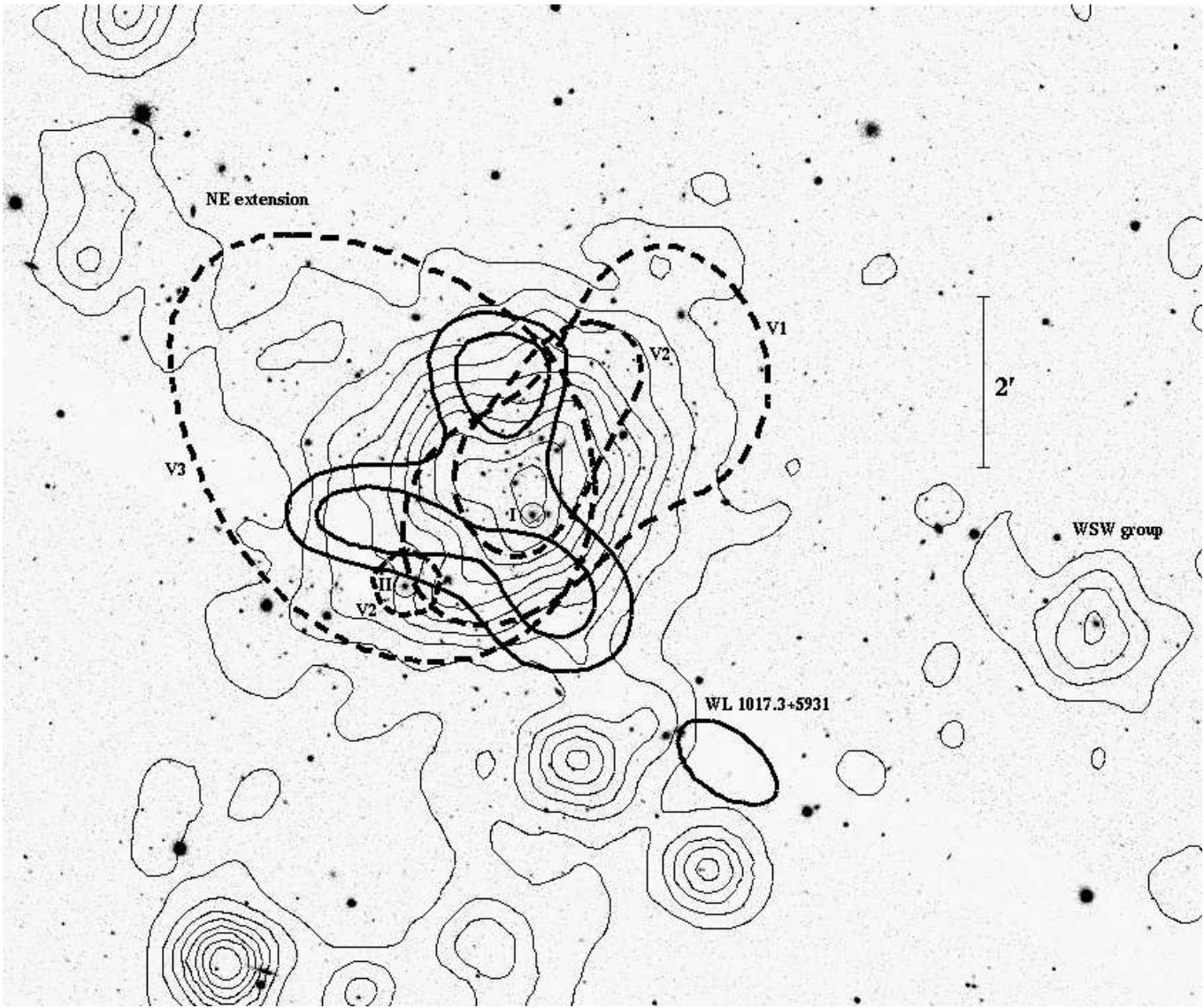}
\caption{Multiwavelength image of A959 (north at the top and east to
  the left). The contour levels (thin levels) of a smoothed archival
  ROSAT/PSPC X--ray image are superimposed on the SDSS $r^\prime$
  image. The peaks (thick contour levels) of the projected mass
  density field (adapted from Dahle et al. \cite{dah03}) reconstructed
  from the weak lensing analysis are shown. Contours recovered from
  the galaxy distributions of the three groups V1, V2, and V3 we detect
  analyzing the velocity field (see text) are also shown (dashed
  thick contours -- one contour for each group). The two black circles
  highlight the first and the second dominant galaxies (I and II, see
  text). Labels indicate individual subclumps discussed in the text.}
\label{figGL}
\end{figure*}

A highly significant mass peak (with some evidence of substructure) is
seen in the weak lensing mass map (Dahle et al. \cite{dah02};
\cite{dah03}), and the mass distribution again indicates two
directions for the cluster elongation (SE--NW and NE--SW) roughly
resembling the galaxy light distribution, although the agreement is
not perfect in the central region (see Fig.~\ref{figGL} for a
multiwavelength view of this cluster). Moreover, there is a dark
matter concentration at $\sim$6\arcm SW of the cluster center, which
does not correspond to any peak in the luminosity distribution and
is connected by a mass bridge with A959 (WL 1017.3+5931).  To date the
nature of this object is unknown and WL 1017.3+5931 remains a good
candidate for an optically ``dark'' subcluster (Dahle et
al. \cite{dah03}).

The presence of a radio extended emission in A959 has been reported by
Cooray et al. (\cite{coo98}). Owen et al. (\cite{owe99}) lists A959
among clusters showing a detectable diffuse radio source with a flux
density of 3 mJy (at 1.4 GHz) and a size of $\sim$3\arcmin ($\sim$0.8
\h in our cosmology). The measurements of this preliminary study are
rough and more observations of this radio source are needed to clarify
the possible connection with the cluster. However, we note that this
source shows the typical sizes of radio halos and radio relics/gischt
(see Kempner et al. \cite{kem04}; Ferrari et
al. \cite{fer08}). Moreover, its radio power ($\sim$1$\times$10$^{24}$
W/Hz) closely matches the relation log$P$(1.4 GHz) -- log$T_{\rm X}$
found by Cassano, Brunetti \& Setti (\cite{cas06}; see their Fig.~3b)
for radio halos.

Poor spectroscopical data exist for A959. The value usually quoted in
the literature for the cluster redshift is $z=0.353$, but Irgens et
al. (\cite{irg02}) determined $z=0.286$ using four galaxies. We
carried out spectroscopic observations at the Telescopio Nazionale
Galileo (TNG) giving new redshift data for 107 galaxies in the field
of A959 and finding $z=0.288$.
 
This paper is organized as follows. We present our new spectroscopical
data and the cluster catalog in Sect.~2.  We present our results about
the cluster structure in Sect.~3.  We briefly discuss our results and
give our conclusions in Sect.~4.

Unless otherwise stated, we give errors at the 68\% confidence level
(hereafter c.l.). Throughout this paper, we use $H_0=70$ km s$^{-1}$
Mpc$^{-1}$ in a flat cosmology with $\Omega_0=0.3$ and
$\Omega_{\Lambda}=0.7$. In the adopted cosmology, 1\arcm corresponds
to $\sim 260$ \kpc at the cluster redshift.

\begin{figure*}
\centering 
\includegraphics[width=18cm]{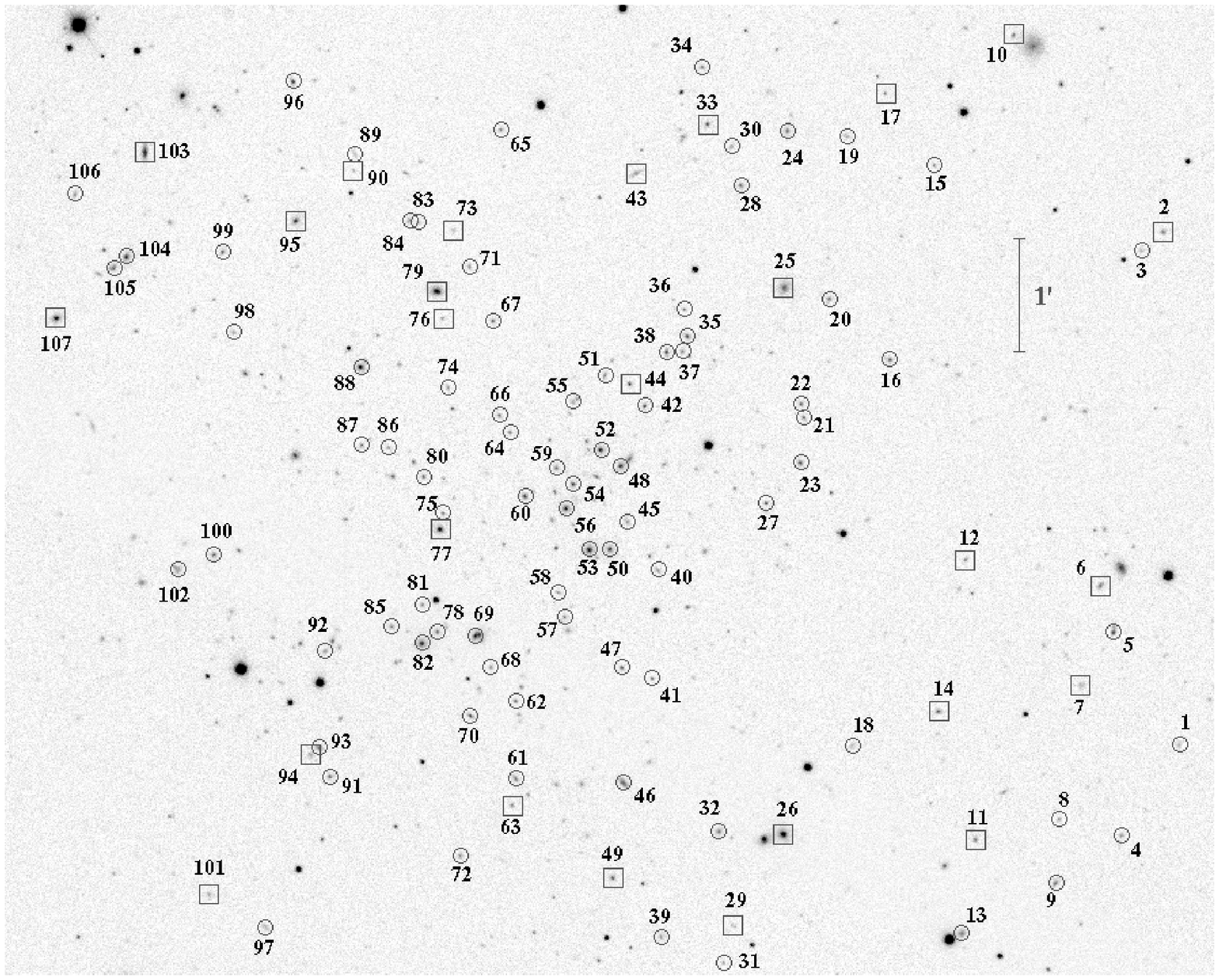}
%\resizebox{\hsize}{!}{\includegraphics{figottico.ps}}
\caption{SDSS $r^\prime$--band image of A959 (north at the top
and east to the left). Circles and boxes indicate cluster members and
non--member galaxies, respectively (see Table~\ref{catalogA959}).}
\label{figottico}
\end{figure*}

\section{New data and galaxy catalog}
\label{newd}

Multi--object spectroscopic observations of A959 were carried out at
the TNG telescope in December 2006. We used DOLORES/MOS with the LR--B
Grism 1, yielding a dispersion of 187 \AA/mm, and the Loral CCD of
$2048\times2048$ pixels (pixel size of 15 $\mu$m). This combination of
grating and detector results in a dispersion of 2.8 \AA/pix. We
observed four MOS masks for a total of 130 slits. We acquired three
exposures of 1800 s for each mask. Wavelength calibration was
performed using helium--argon lamps. Reduction of spectroscopic data
was carried out with the IRAF
\footnote{IRAF is distributed by the National Optical Astronomy
Observatories, which are operated by the Association of Universities
for Research in Astronomy, Inc., under cooperative agreement with the
National Science Foundation.} package.

Radial velocities were determined using the cross--correlation
technique (Tonry \& Davis \cite{ton79}) implemented in the RVSAO
package (developed at the Smithsonian Astrophysical Observatory
Telescope Data Center).  Each spectrum was correlated against six
templates for a variety of galaxy spectral types: E, S0, Sa, Sb, Sc,
Ir (Kennicutt \cite{ken92}). The template producing the highest value
of $\cal R$; i.e., the parameter given by RVSAO and related to the
signal--to--noise of the correlation peak, was chosen. Moreover, all
spectra and their best correlation functions were examined visually to
verify the redshift determination. The median value of $\cal R$ of our
successfully measured galaxy redshifts is $\sim7.5$. In 4 cases (IDs
76, 79, 84, and 90; see Table \ref{catalogA959}) we took the EMSAO
redshift as a reliable estimate of the redshift. Our spectroscopic
survey in the field of A959 consists of 119 spectra with a median
nominal error on $cz$ of 56 \kss.  The nominal errors as given by the
cross--correlation are known to be smaller than the true errors (e.g.,
Malumuth et al. \cite{mal92}; Bardelli et al. \cite{bar94}; Ellingson
\& Yee \cite{ell94}; Quintana et al. \cite{qui00}). Double redshift
determinations for the same galaxy allowed us to estimate real
intrinsic errors in data of the same quality taken with the same
instrument (Barrena et al. \cite{bar07a}, \cite{bar07b}). Here we
applied a similar correction to our nominal errors; i.e., hereafter we
assume that true errors are larger than nominal cross--correlation
errors by a factor 1.4. Thus the median error on $cz$ is $\sim$ 78
\kss.

We also used public photometric data from the Sloan Digital Sky Survey
(SDSS, Data Release 6). In particular, we used $r^\prime$, $i^\prime$,
$z^\prime$ magnitudes, already corrected for the Galactic extinction,
and considered galaxies within a radius of 30\arcm from the cluster
center. Our spectroscopic sample is $\sim$70\% complete down to
$r^\prime=20$ within 4$^{\prime}$ from the cluster center.

Table~\ref{catalogA959} lists the velocity catalog (see also
Fig.~\ref{figottico}): identification number of each galaxy, ID
(Col.~1); right ascension and declination, $\alpha$ and $\delta$
(J2000, Col.~2); SDSS $r^\prime$ magnitudes (Col.~3); heliocentric radial
velocities, ${\rm v}=cz_{\sun}$ (Col.~4) with errors, $\Delta {\rm v}$
(Col.~5).

\input{tab1.tex}

\section{Analysis and results}
\label{anal}

\subsection{Member selection and global properties}
\label{memb}

\begin{figure}
\centering
\resizebox{\hsize}{!}{\includegraphics{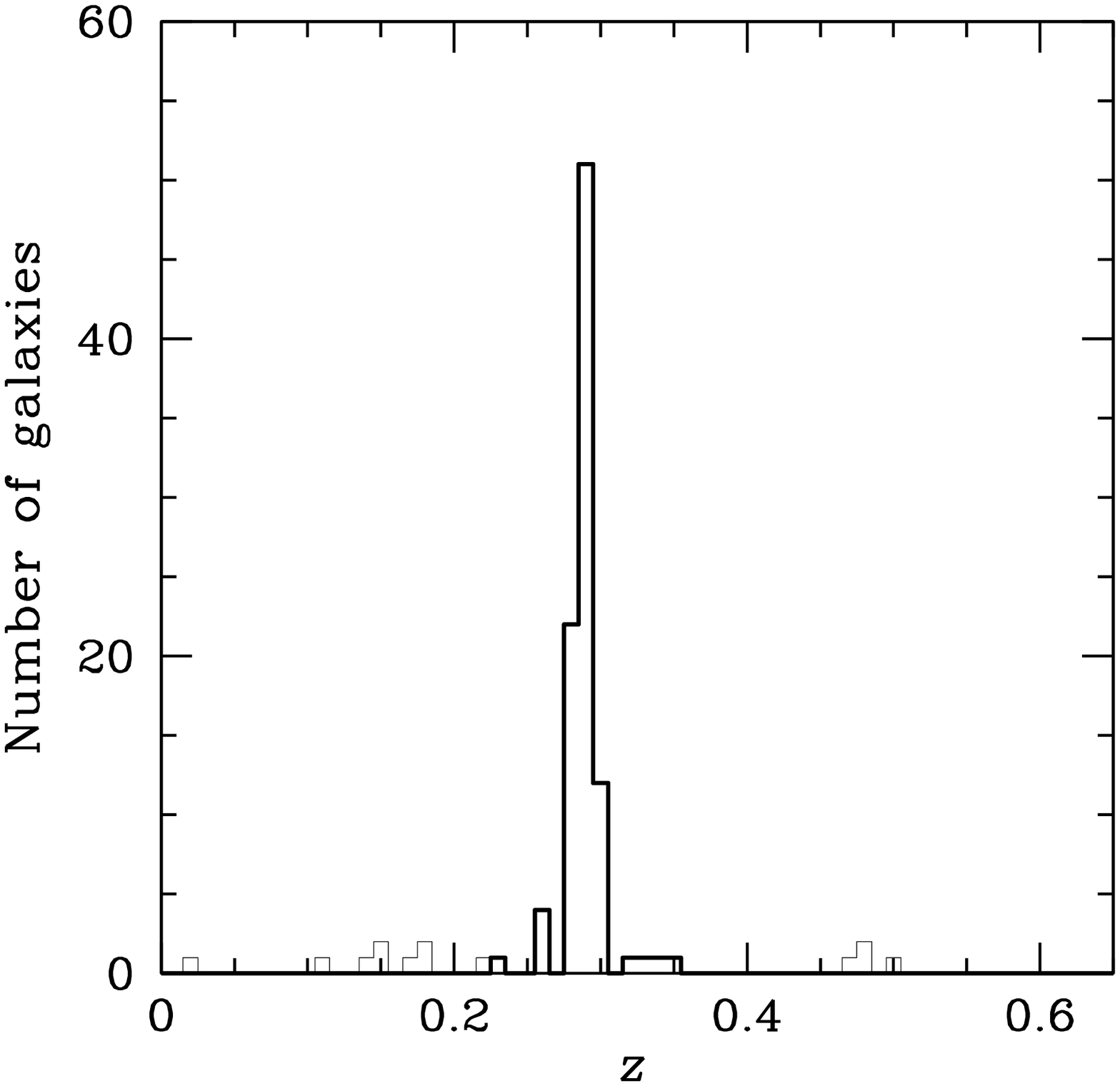}}
\caption
{Redshift galaxy distribution. The solid line histogram refers to
  (94) galaxies assigned to the cluster according to the DEDICA
  reconstruction method.}
\label{fighisto}
\end{figure}

To select cluster members out of 107 galaxies having redshifts, we
follow a two--steps procedure. First, we perform the 1D
adaptive--kernel method (hereafter DEDICA, Pisani \cite{pis93} and
\cite{pis96}; see also Fadda et al. \cite{fad96}; Girardi et
al. \cite{gir96}; Girardi \& Mezzetti \cite{gir01}). We search for
significant peaks in the velocity distribution at $>$99\% c.l.. This
procedure detects A959 as a peak at $z\sim0.288$ populated by 94
galaxies considered as candidate cluster members (see
Fig.~\ref{fighisto}). Out of 13 non members, nine and four are
foreground and background galaxies, respectively.

\begin{figure}
\centering 
\resizebox{\hsize}{!}{\includegraphics{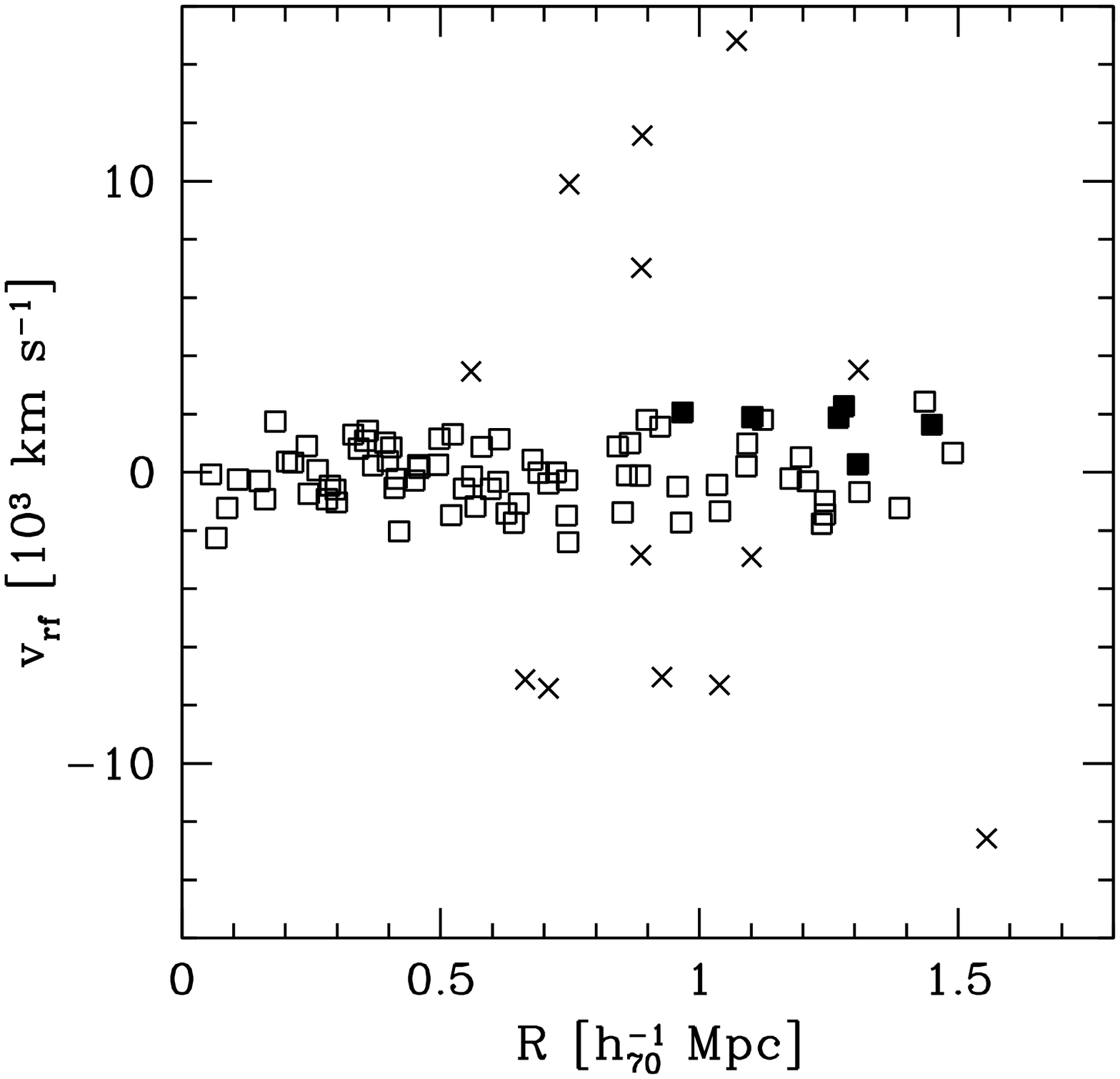}}
\caption
{Rest frame velocity vs. projected clustercentric distance 
  for the 94 galaxies in the main peak (Fig.~\ref{fighisto})
  showing galaxies detected as interlopers by our ``shifting gapper''
  procedure (crosses). Solid squares indicate cluster members detected
  as the northeastern substructure in the Dressler--Shectman test (DS--NE).} 
\label{figvd}
\end{figure}

All the galaxies assigned to the A959 peak are analyzed in the second
step, that uses the combination of position and velocity information:
the ``shifting gapper'' method by Fadda et al. (\cite{fad96}).  This
procedure rejects galaxies that are too far in velocity from the main
body of galaxies and within a fixed bin that shifts along the distance
from the cluster center.  The procedure is iterated until the number
of cluster members converges to a stable value.  Following Fadda et
al. (\cite{fad96}) we use a gap of $1000$ \ks -- in the cluster
rest frame -- and a bin of 0.6 \hh, or large enough to include 15
galaxies. 

The choice of the cluster center is not obvious. No evident dominant
galaxy is present, but rather there are two luminous galaxies (IDs. 53
and 82, hereafter I and II dominant galaxies) defining the SE--NW
direction already shown by X--rays (Dahle et
al. \cite{dah03}). Moreover, the gravitational lensing analysis brings
out several mass peaks (Dahle et al. \cite{dah03}). As the cluster
center, hereafter we assume the position of the peak of X--ray
emission as listed by Popesso et al. (\cite{pop04})
[R.A.=$10^{\mathrm{h}}17^{\mathrm{m}}35\dotsec04$, Dec.=$+59\degree
33\arcmm 27\dotarcs7$ (J2000.0)] which is very close to the I dominant
galaxy. After the ``shifting gapper'' procedure, we obtain a sample of
81 fiducial cluster members (see Fig.~\ref{figvd}).

The 2D galaxy distribution analyzed through the 2D DEDICA method only
shows one peak [at R.A.=$10^{\mathrm{h}}17^{\mathrm{m}}36\dotsec3$,
Dec.=$+59\degree 34\arcmm 09\dotarcs0$ (J2000.0)], which is very close
to the I dominant galaxy (ID.~53), too. Using this alternative cluster
center, we verify the robustness of our member selection obtaining the
same sample of fiducial members as above.

By applying the biweight estimator to the 81 cluster members (Beers et
al. \cite{bee90}), we compute a mean cluster redshift of
$\left<z\right>=0.2883\pm$ 0.0004, i.e.
$\left<\rm{v}\right>=86442\pm$130 \kss.  We estimate the LOS velocity
dispersion, $\sigma_{\rm V}$, by using the biweight estimator and
applying the cosmological correction and the standard correction for
velocity errors (Danese et al. \cite{dan80}).  We obtain $\sigma_{\rm
V}=1170_{-73}^{+83}$ \kss, where errors are estimated through a
bootstrap technique.

\begin{figure}
\centering
\resizebox{\hsize}{!}{\includegraphics{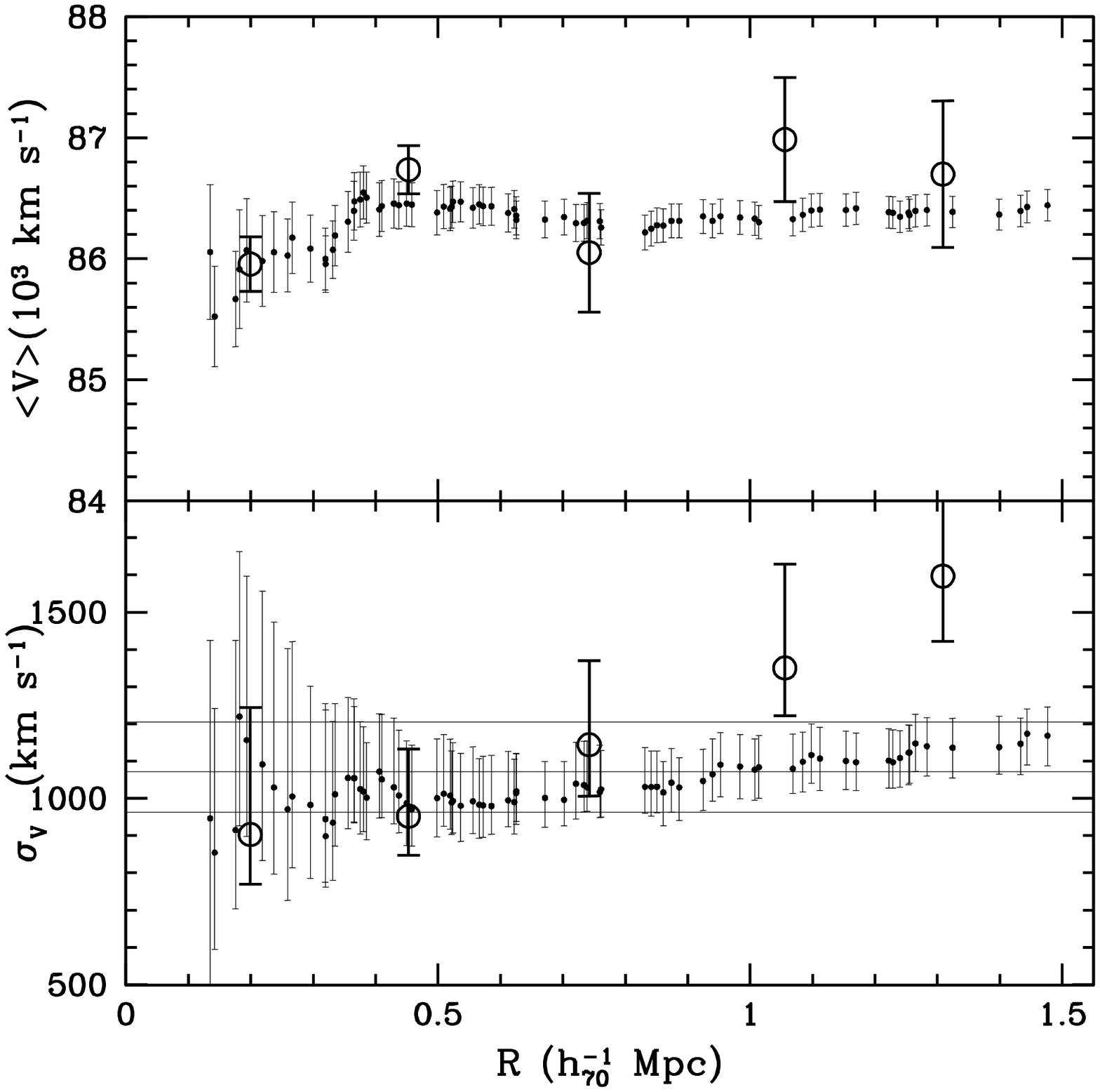}}
\caption
{Differential (big circles) and integral (small points) profiles of
mean velocity ({\em upper panel}) and LOS velocity dispersion ({\em
lower panel}).  For the differential profiles we plot the values for
five annuli from the center of the cluster, each of 0.3 \h (large
symbols).  For the integral profiles, the mean and dispersion at a
given (projected) radius from the clump--center is estimated by
considering all galaxies within that radius -- the first value
computed on the five galaxies closest to the center. The error bands
at the $68\%$ c.l. are also shown.  In the lower panel, the horizontal
line represents the X--ray temperature with the respective 90 per cent
errors (Mushotzky \& Scharf \cite{mus97}) transformed in $\sigma_{\rm
V}$ assuming the density--energy equipartition between gas and
galaxies, i.e.  $\beta_{\rm spec}=1$ (see text).}
\label{figprof}
\end{figure}

To evaluate the robustness of the $\sigma_{\rm V}$ estimate, we
analyze the velocity dispersion profile (Fig.~\ref{figprof}).  The
integral profile is roughly flat suggesting that the value of
$\sigma_{\rm V}$ is quite robust. However we notice a slight
increasing in the external cluster regions confirmed by the behavior
of the differential velocity dispersion profile that increases, rather
than decreases, going to larger and larger radii. The $\sigma_{\rm V}$
enhancement is likely due to the NE structure we detect and discuss in
the following section. A more reliable estimate of the global
$\sigma_{\rm V}$ then seems the value within 0.8 \h where $\sigma_{\rm
V}=1025_{-75}^{+104}$ \kss.

Making the usual assumptions (cluster sphericity, dynamical
equilibrium and galaxy distribution tracing the mass distribution), we
can compute global virial quantities using our estimate of
$\sigma_{\rm V}$. Following the prescriptions of Girardi \& Mezzetti
(\cite{gir01}), we assume for the radius of the quasi--virialized
region $R_{\rm vir}=0.17\times \sigma_{\rm V}/H(z) = 2.15$ \h -- see
their Eq.~1 after introducing the scaling with $H(z)$ (see also Eq.~ 8
of Carlberg et al. \cite{car97} for $R_{200}$). Thus the cluster is
sampled out to a large fraction of $R_{200}$.  We compute the virial
mass (Limber \& Mathews \cite{lim60}; see also, e.g., Girardi et
al. \cite{gir98}):

\begin{equation}
M=3\pi/2 \cdot \sigma_{\rm V}^2 R_{\rm PV}/G-{\rm SPT},
\end{equation}

\noindent where SPT is the surface pressure term correction (The \&
White \cite{the86}), and $R_{\rm PV}$ is a projected radius (equal to
two times the projected harmonic radius). The value of $R_{\rm PV}$
depends on the size of the sampled region and possibly on the quality
of the spatial sampling (e.g., whether the cluster is uniformly
sampled or not). On the whole cluster region sampled out to 1.48 \hh,
we compute $R_{\rm PV}(R<1.48\hhh)=(1.24\pm0.09)$ \hh.  The value of
SPT strongly depends on the radial component of the velocity
dispersion at the radius of the sampled region and could be obtained
by analyzing the velocity--dispersion profile, although this procedure
would require several hundred galaxies.  By combining data on many
clusters sampled out to about $R_{\rm vir}$ a typical value of
SPT$\sim 20\%$ is obtained (Carlberg et al. \cite{car97}; Girardi et
al. \cite{gir98}).  We assume the same correction for A959. Thus, we
obtain $M(<R=1.48 \hhh)=1.15^{-0.19}_{+0.25}$ \mquii.

Since A959 is not sampled out to $R_{\rm vir}$ and the sample is not
magnitude complete we decide to use an alternative estimate of $R_{\rm
  PV}$ based on the knowledge of the galaxy distribution. Following
Girardi et al. \cite{gir98} (see also Girardi \& Mezzetti
\cite{gir01}), we assume a King--like distribution with parameters
typical of nearby/medium--redshift clusters: a core radius $R_{\rm
  c}=1/20\times R_{\rm vir}$ and a slope--parameter $\beta_{\rm
  fit}=0.8$; i.e., the galaxy volume density at large radii goes as
$r^{-3 \beta_{fit}}=r^{-2.4}$. We obtain $R_{\rm PV}(R<2.15\hhh)=1.60$
\hh, where a $25\%$ error is expected (Girardi et
al. \cite{gir98}). We obtain $M(<R_{\rm vir}=2.15
\hhh)=1.47^{+0.47}_{-0.43}$ \mquii, which rescaled using the above
model gives $M(<R=1.48 \hhh)=1.10$ \mquii, in good agreement with the
direct mass estimate given in the previous paragraph.

\subsection{Velocity distribution}
\label{velo}

We analyze the velocity distribution to look for possible deviations
from Gaussianity that might provide important signatures of complex
dynamics. For the following tests the null hypothesis is that the
velocity distribution is a single Gaussian.

We estimate three shape estimators, i.e. the kurtosis, the skewness,
and the scaled tail index (see, e.g., Beers et al.~\cite{bee91}).
According to the value of the kurtosis (2.319), the velocity
distribution is light--tailed and marginally differs from a Gaussian
at the $90-95\%$ c.l. (see Table~2 of Bird \& Beers \cite{bir93}).

\begin{figure}
\centering
\resizebox{\hsize}{!}{\includegraphics{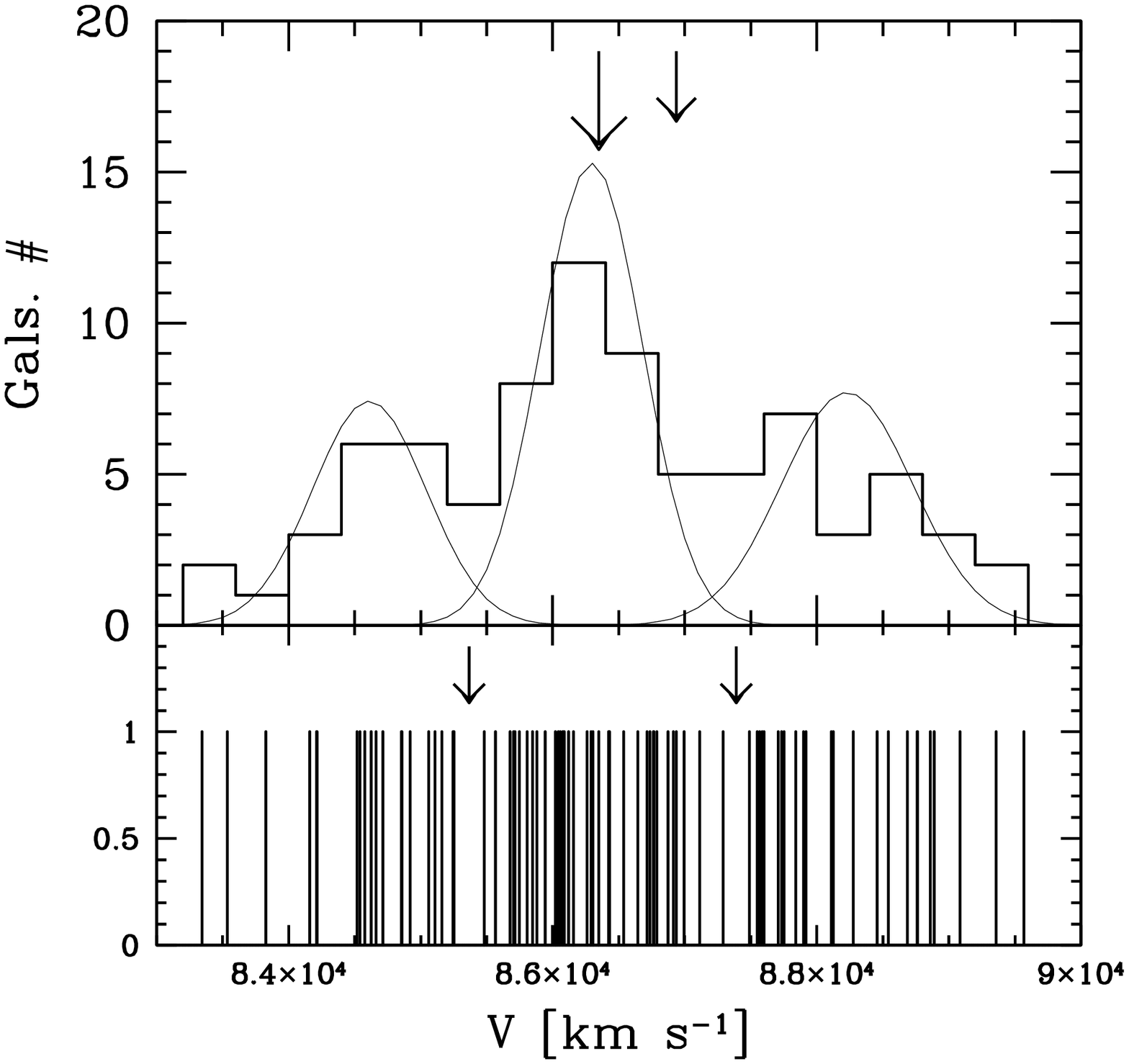}}
\caption
{{\em Upper panel}: rest frame velocity histogram for the 81 cluster
  members. The bigger and smaller arrows indicate the velocities of
  the I and II dominant galaxies. The three Gaussians correspond
  to the three groups (V1, V2, and V3, from left to right) found in the
  velocity distribution (see Table~2). {\em Lower panel}: stripe
  density plot where the arrows indicate the positions of the
  significant gaps.}
\label{figstrip}
\end{figure}

Then we investigate the presence of gaps in the velocity distribution.
A weighted gap in the space of the ordered velocities is defined as
the difference between two contiguous velocities, weighted by the
location of these velocities with respect to the middle of the
data. We obtain values for these gaps relative to their average size,
precisely the midmean of the weighted--gap distribution. We look for
normalized gaps larger than 2.25, since in random draws of a Gaussian
distribution they arise at most in about $3\%$ of the cases,
independent of the sample size (Wainer and Schacht~\cite{wai78}; see
also Beers et al.~\cite{bee91}). We detect two significant gaps (at
the $97\%$ c.l.) that separate the main cluster in three groups of 20,
37 and 24 galaxies (see Fig.~\ref{figstrip}), hereafter named V1, V2,
and V3, respectively (see Table~\ref{tabv} for their main kinematical
properties).

\input{tab2.tex}

\begin{figure}
\centering 
\resizebox{\hsize}{!}{\includegraphics{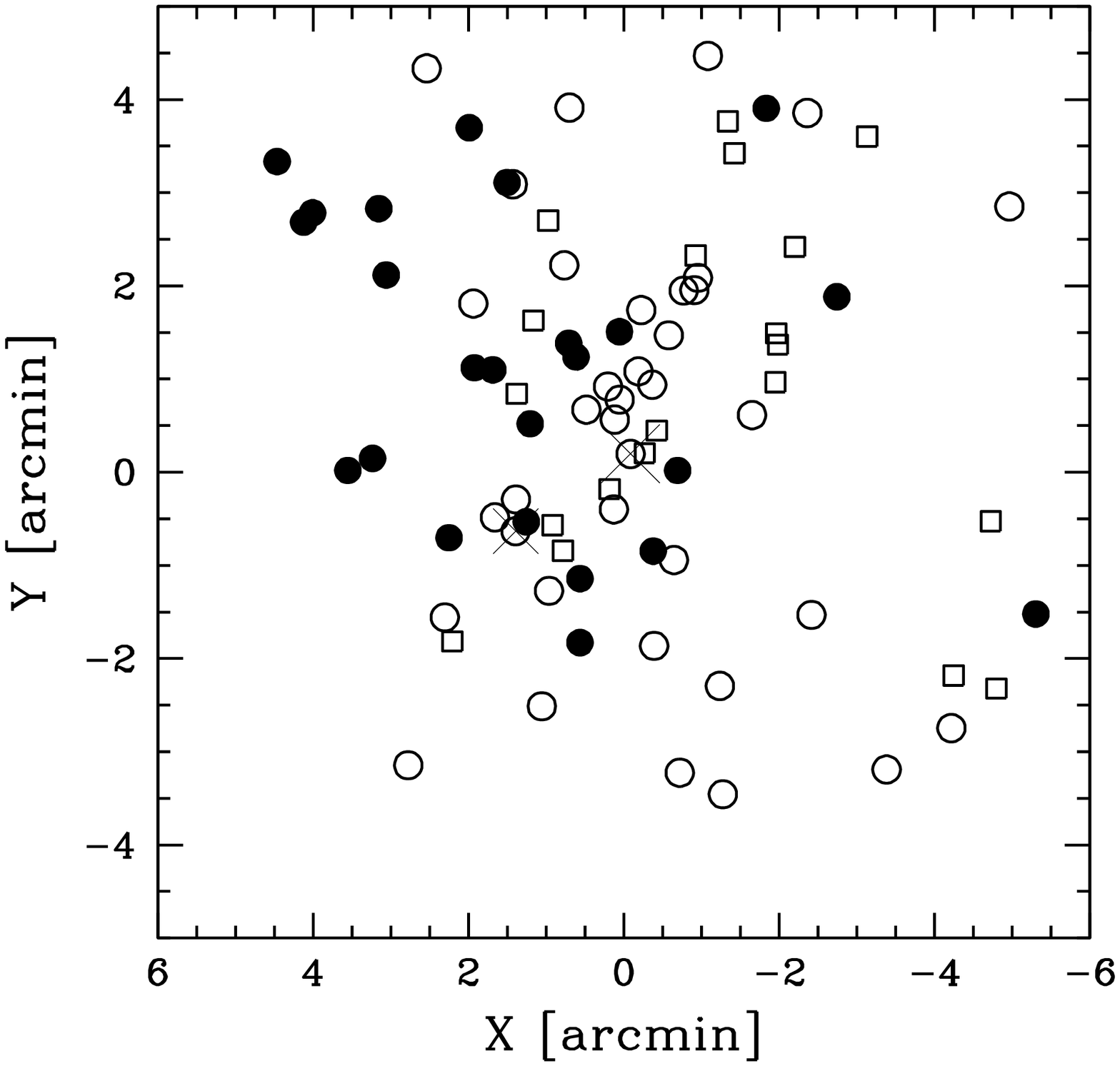}}
\caption
{Spatial distribution on the sky of the 81 member galaxies showing the
  groups recovered by the weighted gap analysis. Open squares, open
  circles and solid circles indicate V1, V2 and V3 groups. The larger
  and smaller crosses indicate the positions of the I and II dominant
  galaxies, respectively.}
\label{figv123}
\end{figure}

When compared two by two through the 2D Kolmogorov--Smirnov test
(Fasano \& Franceschini \cite{fas87}), these groups differ in spatial
distribution: the V1 group marginally differs both from the V2 and V3
groups (at the $90\%$ and $94\%$ c.l., respectively); the V2 group
differs from the V3 group (at the $98\%$ c.l.). In particular, the V2
group forms a dense elongated group around the cluster center and
hosts both dominant galaxies. The V3 galaxies are mostly located in
the northeastern part of the sampled region. For each of the these
groups, Fig.~\ref{figv123} shows the spatial distribution of
galaxies. The contours of the three galaxy distributions as recovered
using the 2D DEDICA method (one for each distribution) are shown in
Fig.~\ref{figGL}. The V1 and V2 groups define the SE--NW direction
and, in particular, the V2 group shows a bimodal 2D distribution where
the two peaks correspond to the two dominant galaxies. The V3 group
defines the NE--SW direction.

Finally, we notice that the luminous galaxies are preferentially
located in the V2 group tracing, in particular, the cluster core.
When comparing the magnitude distribution of V2 galaxies with that of
V1+V3 galaxies, we find a difference at the $98\%$ c.l. according to
the 1D Kolmogorov--Smirnov test (see, e.g., Ledermann \cite{led82}).

\subsection{``3D'' substructure analysis}
\label{comb}

The existence of correlations between positions and velocities of
cluster galaxies is a footprint of real substructures.  Here we use
different approaches to analyze the structure of A959 combining
velocity and position information.

\begin{figure}
\centering
\resizebox{\hsize}{!}{\includegraphics{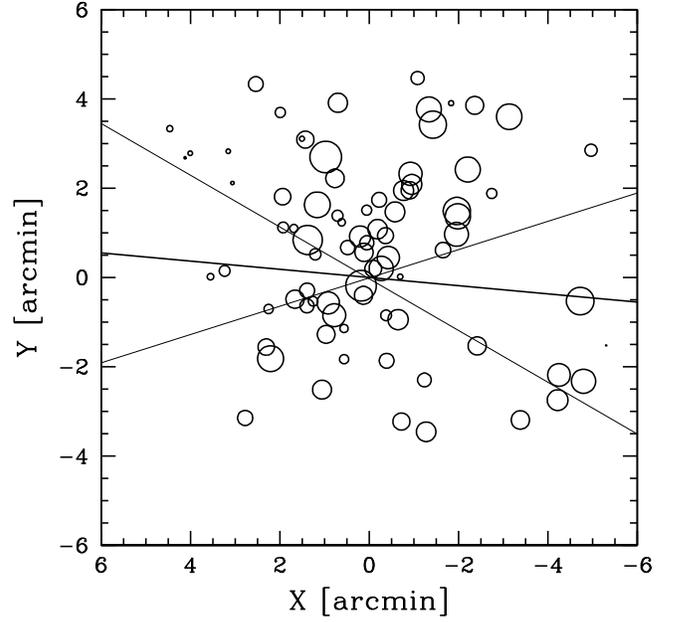}}
\caption
{Spatial distribution on the sky of 81 cluster members.  The larger
the symbol, the smaller the radial velocity.  The plot is centered on
the cluster center.  The solid and faint lines indicate the position
angle of the cluster gradient and relative errors, respectively.}
\label{figgrad}
\end{figure}

We analyze the presence of a velocity gradient performing a multiple
linear regression fit to the observed velocities with respect to the
galaxy positions in the plane of the sky (e.g, Boschin et
al. \cite{bos04} and refs. therein). We find a position angle on the
celestial sphere of $PA=85_{-20}^{+23}$ degrees (measured
counter--clockwise from north), i.e. higher--velocity galaxies lie
in the western region of the cluster (see Fig.~\ref{figgrad}). To
assess the significance of this velocity gradient, we perform 1000
Monte Carlo simulations by randomly shuffling the galaxy velocities,
and for each simulation we determine the coefficient of multiple
determination ($RC^2$, see e.g., NAG Fortran Workstation Handbook
\cite{nag86}).  We define the significance of the velocity gradient as
the fraction of times in which the $RC^2$ of the simulated data is
smaller than the observed $RC^2$. We find that the velocity gradient
is significant at the $99.6\%$ c.l..

\begin{figure}
\centering 
\resizebox{\hsize}{!}{\includegraphics{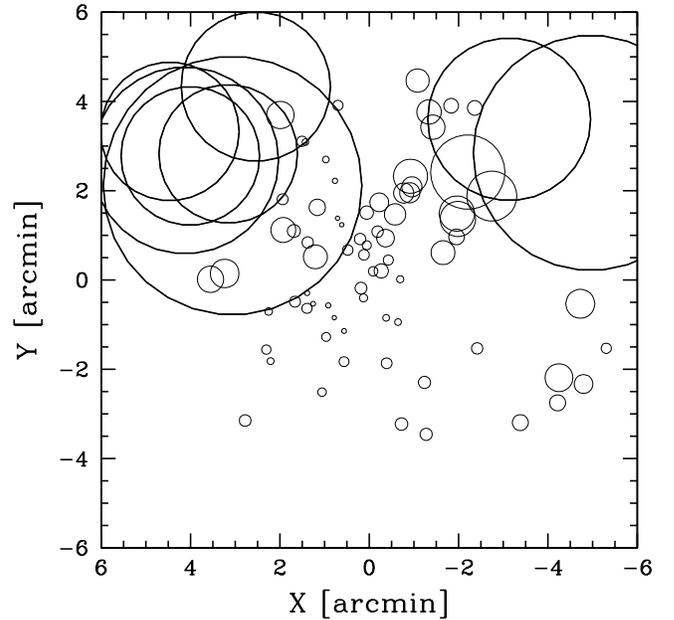}}
\caption
{Spatial distribution of the 81 cluster members, each marked by a
circle: the larger the circle, the larger the deviation $\delta_i$ of
the local parameters from the global cluster parameters; i.e., there is
more evidence of substructure according to the Dressler \& Shectman
test (see text). The boldface circles indicate those with $\delta_i
\ge 3.5$ (see text). Out of these, the six northern galaxies form the
DS-NE clump. The plot is centered on the cluster center.}
\label{figds}
\end{figure}

We combine galaxy velocity and position information to compute the
$\Delta$--statistics devised by Dressler \& Shectman (\cite{dre88}).
This test is sensitive to spatially compact subsystems that have
either an average velocity that differs from the cluster mean or a
velocity dispersion that differs from the global one, or both.  We
find $\Delta=126$ for the value of the parameter that gives the
cumulative deviation.  This value is a significant indication of
substructure (at the $99.9\%$ c.l.) as assessed by computing 1000
Monte Carlo simulations, randomly shuffling the galaxy velocities.
Figure~\ref{figds} shows the distribution on the sky of all galaxies,
each marked by a circle: the larger the circle, the larger the
deviation $\delta_i$ of the local parameters from the global cluster
parameters, i.e. the greater the evidence for substructure.

\begin{figure}
\centering 
\resizebox{\hsize}{!}{\includegraphics{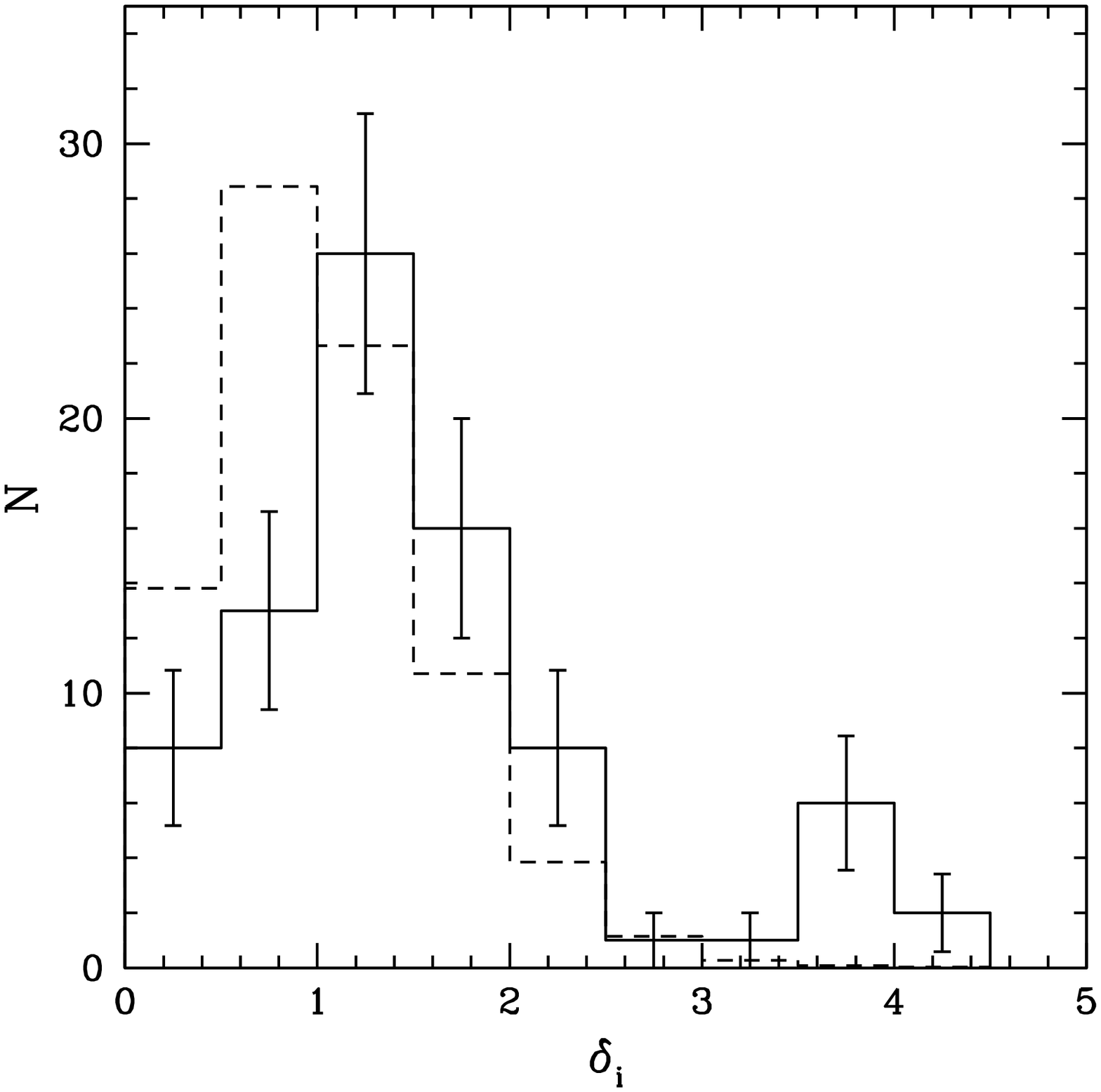}}
\caption
{The distribution of $\delta_i$ deviations of the Dressler--Shectman
  analysis for the 81 member galaxies. The solid line represents the
  observations, the dashed line the distribution for the galaxies of
  simulated clusters, normalized to the observed number.}
\label{figdeltai}
\end{figure}

To better point out galaxies belonging to substructures, we resort to
the technique developed by Biviano et al. (\cite{biv02}, see also
Boschin et al. \cite{bos06}; Girardi et al. \cite{gir06}), who used
the individual $\delta_i$--values of the Dressler--Shectman
method. The critical point is to determine the value of $\delta_i$
that optimally indicates galaxies belonging to substructure. To this
aim we consider the $\delta_i$--values of all 1000 Monte Carlo
simulations used above.  The resulting distribution of $\delta_i$ is
compared with the observed one finding a difference at the $99.6\%$
c.l. according to the 1D Kolmogorov--Smirnov test.  The ``simulated''
distribution is normalized to produce the observed number of galaxies
and compared with the observed distribution in Fig.~\ref{figdeltai}:
the latter shows a tail at high values. The inspection of
Fig.~\ref{figdeltai} suggests that galaxies with $\delta_i > 3.5$
likely are in substructures. These galaxies are indicated with heavy
circles in Fig.~\ref{figds}. The six galaxies located in the
northeastern region of the cluster are likely to form a subclump,
hereafter DS--NE (see Table~\ref{tabv} and Fig.~\ref{figvd}).

\begin{figure}
\centering 
\resizebox{\hsize}{!}{\includegraphics{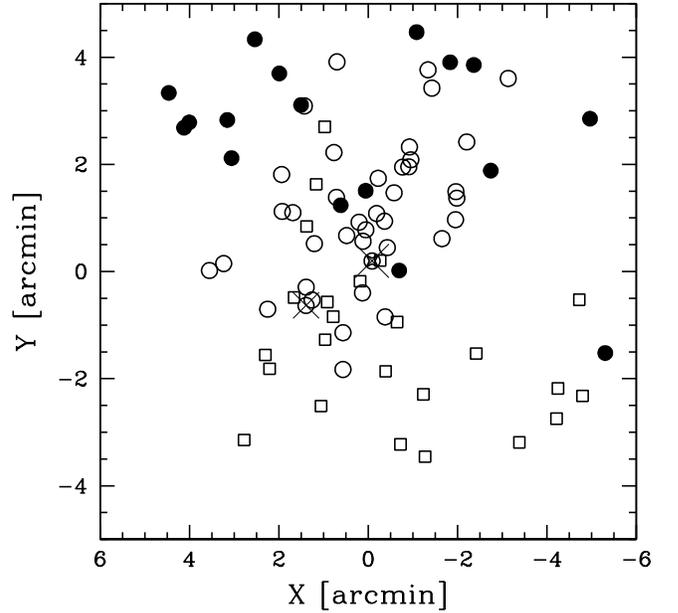}}
\caption
{Spatial distribution on the sky of the 81 galaxies
  of the whole cluster showing the groups recovered by the weighted
  gap analysis. Open squares, open circles, and solid circles indicate
  KMM1, KMM2, and KMM3 groups. The larger and smaller crosses
  indicate the positions of the I and II dominant galaxies,
  respectively.}
\label{figkmm}
\end{figure}

We also use here the Kaye's mixture model (KMM) test to find a
possible group partition of the velocity distribution (as implemented
by Ashman et al. \cite{ash94}). The KMM algorithm fits a
user--specified number of Gaussian distributions to a dataset and
assesses the improvement of that fit over a single Gaussian. In
addition, it provides the maximum--likelihood estimate of the unknown
n--mode Gaussians and an assignment of objects into groups. This
algorithm is usually used to analyze the velocity distribution where
theoretical and/or empirical arguments indicate that the Gaussian
model is reasonable. Here the 1D KMM test fails in confirming the
three groups detected by the weighted gap method. However, since in
this case there is a significant correlation between galaxy velocities
and positions we have decided to use KMM as a 3D diagnostic. We use
the three--group galaxy assignment given by the weighted gap analysis
to determine the first guess when fitting three groups. The algorithm
fits a three--group partition at the $92\%$ c.l. according to the
likelihood ratio test (hereafter KMM1, KMM2, KMM3 groups of 24, 40,
and 17 galaxies from low to high mean velocities). The results for the
three groups are shown in Table~\ref{tabv} and Fig.~\ref{figkmm}.
Although there are some differences between the group memberships
recovered with the two different methods (cf. Fig.~\ref{figkmm} with
Fig.~\ref{figv123}), there are important common features: the NE
region is populated by the high--velocity galaxies while the cluster
core, elongated in the SE--NW direction, is formed by the medium
velocity galaxies. The main difference is that the SE--NW bimodal
structure is now somehow split, with KMM1 and KMM2 tracing the SE and
NW peaks better, respectively. However, note that the two dominant
galaxies are both contained in the medium velocity group
KMM2. Moreover, the galaxies in SW region are now assigned primarily
to the low velocity group.

\subsection{2D galaxy distribution}
\label{sec:2D}

When applying the DEDICA method to the 2D distribution of the 81
galaxy members, we find only one significant peak.  However, our
spectroscopic data suffer from magnitude incompleteness, and the
sampled region does not cover the whole virial region. To overcome
these limits we use the photometric catalog extracted from the SDSS
(Data Release 6).

\begin{figure}
%\centering
\includegraphics[width=8cm]{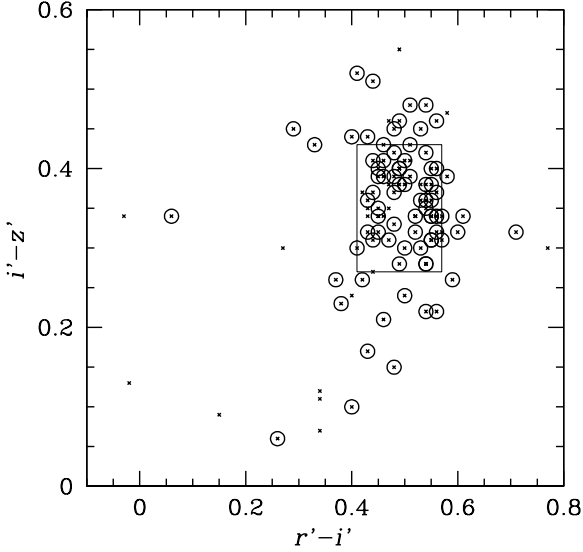}
\caption
{$i'$--$z'$ vs. $r'$-$i'$ diagram for galaxies with available
spectroscopy is shown by small crosses. The faint box is centered
on the median value for colors of member galaxies (circles) and
encloses galaxies having colors in the range of $\pm$0.08 mag from
median values.}
\label{figcc}
\end{figure}

\begin{figure}
%\centering
\includegraphics[width=8cm]{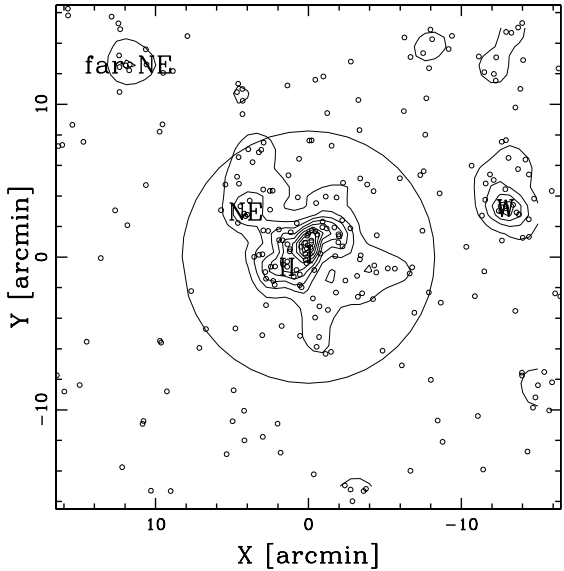}
\caption
{Spatial distribution on the sky and relative isodensity contour map
  of likely cluster members (according to the selection in the
  color--color plane) with $r'\le 21$, obtained with the DEDICA
  method. Dominant galaxies I and II, the NE, W, and far NE peaks are
  indicated. The circle indicates the likely virialized region.}
\label{figk2cc}
\end{figure}

In the SDSS photometric catalog, we select likely members on the base
of the ($r'$--$i'$) -- ($i'$--$z'$) color--color plane (see
e.g. Goto et al. \cite{got02}; Boschin et al. \cite{bos08}). Out of
our photometric catalog, we consider galaxies (here objects with $r'$
phototype parameter $=3$) lying within $\pm$0.08 mag from the median
values of $r'$--$i'$=0.49 and $i'$--$z'$=0.35 colors of the
spectroscopically cluster members (see Fig.~\ref{figcc}). The value of
0.08 mag is two times the typical scatter reported by Goto et
al. (\cite{got02}) for the corresponding color--magnitude relations
$r'$--$i'$ vs. $r'$ and $i'$--$z'$ vs. $r'$. By adopting this
alternative selection method in a region within 30\arcm (almost 4
$R_{\rm vir}$) from the cluster center we select a sample of likely
cluster members with $r'<21$. The result of the application of DEDICA
to this sample is shown in Fig.~\ref{figk2cc}. The galaxy distribution
shows a shape elongated along the SE--NW direction in the central
regions. The most significant peaks (at the $>99.9\%$ c.l.)
correspond to the I dominant galaxy, the II dominant galaxy, a western
peak well outside of $R_{\rm vir}$ (hereafter ``W'' peak), the NE peak
at $\lesssim$5\arcmin and a far NE peak at $\sim$18\arcm from the
cluster center (hereafter ``far NE'' peak).

\begin{figure}
%\centering
\includegraphics[width=9cm]{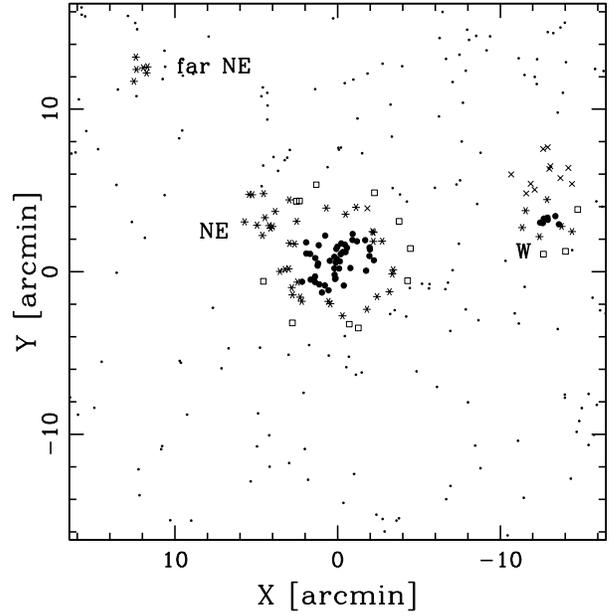}
\caption
{Galaxies belonging to structures as detected by the Voronoi
tessellation and percolation technique. The algorithm is run on a
sample of galaxies with $r'\le 22$ selected in the color--color
plane (see text) within 30\arcm from the center of A959. Open
squares, crosses, asterisks, and solid circles indicate galaxies in
structures at the 95$\%$, 98$\%$, 99$\%$, and 99.9$\%$ c.ls.,
respectively. The NE clump and the W and far NE groups are
indicated.}
\label{figvtp}
\end{figure}

To probe the robustness of these detections, we also apply the Voronoi
tessellation and percolation (VTP) technique (see Ramella et
al. \cite{ram01}; Barrena et al. \cite{bar05}). This technique is
non--parametric and does not smooth the data. As a consequence, it
identifies galaxy structures irrespective of their shapes. In
particular, we run VTP in a region within 30\arcm from the cluster
center selecting galaxies with $r'<22$ and the same color constraints
as above. The result of the application of VTP is shown in
Fig.~\ref{figvtp}. VTP is run 4 times adopting four detection
thresholds: galaxies identified as belonging to structures at the
95$\%$, 98$\%$, 99$\%$, and $99.9\%$ c.ls. VTP reveals the elongation
of A959 in the SE--NW direction and confirms the NE peak, the W, and
far NE peaks.

\section{Discussion and conclusions}
\label{concl}

Our analysis based on member galaxies shows that A959 is a strongly
substructured cluster. In particular, we detect an NE clump, at
$\sim$5\arcm from the cluster center, using the weighted gap analysis
(V3 group), the Dressler--Shectman statistics (DS--NE group), the 3D
KMM analysis (KMM3 group) and the 2D analysis of the galaxy
distribution.  This clump has a rest frame velocity of v$_{\rm
rf}\sim +1900$ \ks and is the likely cause of the velocity gradient we
observe in A959. The NE clump is also detected in ROSAT/PSPC X--ray
data and in the lensing gravitational map by Dahle et
al. (\cite{dah03}). This clump is also aligned with the SW mass
concentration WL 1017.3+5931 whose nature is very intriguing.

The central cluster region is not well relaxed, too.  We find evidence
of a central, dense structure elongated along the SE--NW direction as
shown by the the weighted gap analysis (V2 group), the 3D KMM analysis
(KMM2 group) and the 2D analysis of the galaxy distribution. The
SE--NW central structure has an intermediate velocity and hosts more
luminous cluster galaxies; thus it likely represents the main part of
the cluster still showing the trace of a past merger. The apparent
elongation of this structure is connected with the presence of the two
dominant galaxies and some galaxies surrounding them, as shown by the
two peaks detected by DEDICA and VTP in the 2D galaxy
distributions. Close couples of dominant galaxies having different
velocities are often observed in clusters (Boschin et
al. \cite{bos06}; Barrena et al. \cite{bar07a}) and are the likely
tracers of a previous cluster merger. Indeed cluster merger is thought
to be the cause of the formation of dumbbell galaxies (e.g., Beers et
al. \cite{bee92}; Flores et al. \cite{flo00}). With respect to other
cases we have studied (e.g., clusters Abell 773 and Abell 2744; see
Barrena et al. \cite{bar07a} and Boschin et al. \cite{bos06},
respectively), the two dominant galaxies are closer in the velocity
space and more distant in sky positions suggesting that the direction
of the old, past merger is almost parallel to the plane of the sky.

\input{tab3.tex}

The SE--NW elongation is also shown by gravitational lensing and ROSAT
X--ray data (Dahle et al. \cite{dah03}). The ellipticity of the X--ray
surface brightness appears less pronounced than those of mass and
galaxy distributions, in agreement with the collisional component of
the cluster having already had some time to relax.

The above discussion suggests that A959 is forming along two main
directions of mass accretion (SE--NW and NE--SW). Indeed, the region
around this cluster is quite complex and deserves a brief
discussion. By running the VTP technique and DEDICA on the 2D galaxy
distribution we find two significant galaxy concentrations: a group at
$\sim$13\arcm W of A959 and a far NE one at $\sim$18\arcm from the
core of the cluster. These two groups were first detected by Koester
et al. (\cite{koe07}) using the maxBCG red--sequence method in SDSS
photometric data. They computed photometric redshifts of $\sim$0.281
and $\sim$0.284 for the W and NE groups, respectively. Thus, they are
at the same redshift of A959, at projected distances from it of
$\sim$3.4 \h (W group) and $\sim$4.7 \h (NE group), possibly
representing the most important knots of the filamentary large--scale
structure around the cluster, after A959 itself. Closer to A959 (at
$\sim$6\arcm WSW from the cluster center), ROSAT archival data reveal
a diffuse X--ray emission probably associated to a small galaxy
group. Unfortunately, we have no redshift in this region, but at the
center of the X--ray emission there is a brilliant galaxy (only
slightly less luminous than the I and II dominant galaxies) showing
the typical colors of the early--type galaxies of A959. This galaxy
(known as NVSS J101642+592222) is an evident pointlike radio source in
NVSS (Condon et al. \cite{con98}) archival data, too. We suspect that
this radio galaxy is at the center of a galaxy group falling in the
gravitational well of A959. In Table~\ref{tabstruct} we summarize the
properties of the WSW group, the W group and other structures not
listed in Table~\ref{tabv}.

As for the global properties of the cluster, the value we find for the
velocity dispersion of all A959 cluster members is very high,
$\sigma_{\rm V}=1170_{-73}^{+83}$ \kss. When considering a more
central region to avoid the NE subclump we still obtain a high value
$\sigma_{\rm V}(<R=0.8\hhh)=1025_{-75}^{+104}$ \kss. This value is in
good agreement with the average X--ray temperature (Mushotzky \&
Scharf \cite{mus97}) when assuming the energy density equipartition
between gas and galaxies (i.e. $\beta_{\rm
spec}=1$\footnote{$\beta_{\rm spec}=\sigma_{\rm V}^2/(kT/\mu m_{\rm
p})$ with $\mu=0.58$ the mean molecular weight and $m_{\rm p}$ the
proton mass.}, see Fig.\ref{figprof}) and leads to a virial mass of
$M(<R=1.48 \hhh)=1.15^{+0.25}_{-0.19}$ \mquii.  We compare our mass
estimate with that derived from the gravitational lensing
analysis. Dahle et al. (\cite{dah02}) obtained a projected mass
$M_{\rm proj}(<1.9\hhh) = 1.6-3.2$ \mqui (in our cosmology). To
rescale and project our mass, we assume two alternative mass
distributions: a King--like mass distribution (see above) or an NFW
profile where the mass--dependent concentration parameter $c$ is taken
from Navarro et al. (\cite{nav97}) and rescaled by the factor $1+z$
(Bullock et al. \cite{bul01}; Dolag et al. \cite{dol04}). Our
projected mass estimate is $M_{\rm proj}(<1.9 \hhh)=(1.5-1.9)$ \mquii,
where the projection is done assuming that the cluster mass
distribution is truncated at one or at two virial radii. The lensing
mass is consistent with our projected estimate within their errors,
even if the former is likely to be a bit higher, possibly due to the
young dynamical state of the cluster.

In conclusion, A959 shows the typical features of most clusters
hosting halo/relic radio sources, i.e. a high mass (higher than
$0.7$ \mqui within 2\hh; see Giovannini \& Feretti \cite{gio02}) and
a young dynamical state (diffuse radio sources appear in
clusters possessing substructures, i.e. experiencing large departures
from a virialized state; see, e.g., Buote \cite{buo01}). This makes
of this cluster a good candidate for deep radio studies. More
observations of the diffuse radio source will allow to clarify its
connection with the internal dynamics of this cluster. Moreover, deep
X--ray observations will be useful for determining the merging phase of
the NE clump.

\begin{acknowledgements}
This publication is based on observations made on the island of La
Palma with the Italian Telescopio Nazionale Galileo (TNG), operated by
the Fundaci\'on Galileo Galilei -- INAF (Istituto Nazionale di
Astrofisica), in the Spanish Observatorio of the Roque
de Los Muchachos of the Instituto de Astrofisica de Canarias.

This research has made use of the NASA/IPAC Extragalactic Database
(NED), which is operated by the Jet Propulsion Laboratory, California
Institute of Technology, under contract with the National Aeronautics
and Space Administration.

This research has made use of the galaxy catalog of the Sloan Digital
Sky Survey (SDSS). Funding for the SDSS has been provided by the
Alfred P. Sloan Foundation, the Participating Institutions, the
National Aeronautics and Space Administration, the National Science
Foundation, the U.S. Department of Energy, the Japanese
Monbukagakusho, and the Max Planck Society. The SDSS Web site is
http://www.sdss.org/.

The SDSS is managed by the Astrophysical Research Consortium for the
Participating Institutions. The Participating Institutions are the
American Museum of Natural History, Astrophysical Institute Potsdam,
University of Basel, University of Cambridge, Case Western Reserve
University, University of Chicago, Drexel University, Fermilab, the
Institute for Advanced Study, the Japan Participation Group, Johns
Hopkins University, the Joint Institute for Nuclear Astrophysics, the
Kavli Institute for Particle Astrophysics and Cosmology, the Korean
Scientist Group, the Chinese Academy of Sciences (LAMOST), Los Alamos
National Laboratory, the Max-Planck-Institute for Astronomy (MPIA),
the Max-Planck-Institute for Astrophysics (MPA), New Mexico State
University, Ohio State University, University of Pittsburgh,
University of Portsmouth, Princeton University, the United States
Naval Observatory, and the University of Washington.

This work was partially supported by a grant from the Istituto
Nazionale di Astrofisica (INAF, grant PRIN--INAF2006 CRA ref number
1.06.09.06).

\end{acknowledgements}

\end{document}

%% file: tab1.tex
%\documentclass[referee]{aa}
%\usepackage{graphicx}
%%new commands
%\def\lesssim{\mathrel{\hbox{\rlap{\hbox{\lower4pt\hbox{$\sim$}}}\hbox{$<$}}}}
%\def\gtrsim{\mathrel{\hbox{\rlap{\hbox{\lower4pt\hbox{$\sim$}}}\hbox{$>$}}}}
%\newcommand{\mincir}{\raise -2.truept\hbox{\rlap{\hbox{$\sim$}}\raise5.truept
%\hbox{$<$}\ }}
%\newcommand{\magcir}{\raise -2.truept\hbox{\rlap{\hbox{$\sim$}}\raise5.truept
%\hbox{$>$}\ }}
%\newcommand{\siml}{\raise -2.truept\hbox{\rlap{\hbox{$\sim$}}\raise5.truept
%\hbox{$<$}\ }}
%\newcommand{\simg}{\raise -2.truept\hbox{\rlap{\hbox{$\sim$}}\raise5.truept
%\hbox{$>$}\ }}
%\newcommand{\be}{\begin{equation}}
%\newcommand{\ee}{\end{equation}}
%\newcommand{\ba}{\begin{eqnarray}}
%\newcommand{\ea}{\end{eqnarray}}
%\newcommand {\h} {$h^{-1}$ Mpc $ \;$}
%\newcommand {\kpc} {$h^{-1}$ kpc}
%\newcommand {\hh} {$h^{-1}$ Mpc}
%\newcommand {\ks} {km~s$^{-1} \;$}
%\newcommand {\kss} {km~s$^{-1}$}
%\newcommand {\mpc} {$Mpc \;$}
%\newcommand {\msun} {$h^{-1} \  M_{\odot} \;$}
%\newcommand {\m} {$M_{\odot} \;$}
%\newcommand {\ml} {$h \, M_{\odot}/L_{\odot} \;$}
%\newcommand {\mll} {$h \, M_{\odot}/L_{\odot}$}
%\newcommand{\vel}{\,{\rm km\,s^{-1}}}
%%
%\begin{document}

%\addtocounter{table}{-2}
\begin{table}[!ht]
        \caption[]{Velocity catalog of 107 spectroscopically measured
galaxies in the field of A959.}
         \label{catalogA959}
              $$ 
        % \begin{array}{p{0.5\linewidth}l}
           \begin{array}{r c c r r}
            \hline
            \noalign{\smallskip}
            \hline
            \noalign{\smallskip}

\mathrm{ID^{\mathrm{a}}} & \mathrm{\alpha},\mathrm{\delta}\,(\mathrm{J}2000)  & r^\prime & \mathrm{v}\,\,\,\,\,\,\, & \mathrm{\Delta}\mathrm{v} \\
  & & &\mathrm{(\,km}&\mathrm{s^{-1}\,)}\\
            \hline
            \noalign{\smallskip}  

  1 &10\ 16\ 53.19 ,+59\ 31\ 56.0&     20.07&  89567& 111\\
\textit{2} &10\ 16\ 54.24 ,+59\ 36\ 28.3&     19.64&  70237&  76\\
  3 &10\ 16\ 55.79 ,+59\ 36\ 18.5&     20.21&  87292&  74\\
  4 &10\ 16\ 57.21 ,+59\ 31\ 08.0&     20.27&  84858&  78\\
  5 &10\ 16\ 57.75 ,+59\ 32\ 55.8&     18.81&  84159&  62\\
\textit{6} &10\ 16\ 58.65 ,+59\ 33\ 20.6&     19.33&  64557&  84\\
\textit{7} &10\ 17\ 00.01 ,+59\ 32\ 27.6&     19.69& 142791&  57\\
  8 &10\ 17\ 01.54 ,+59\ 31\ 16.5&     20.32&  85162&  88\\
  9 &10\ 17\ 01.79 ,+59\ 30\ 42.7&     19.07&  85568& 101\\
\textit{10} &10\ 17\ 04.69 ,+59\ 38\ 13.0&     19.70&  54207& 116\\
 \textit{11} &10\ 17\ 07.36 ,+59\ 31\ 05.8&     19.57&  82691&  53\\
 \textit{12} &10\ 17\ 08.06 ,+59\ 33\ 34.6&     20.08& 101333&  66\\
 13 &10\ 17\ 08.37 ,+59\ 30\ 16.1&     19.08&  86044&  57\\
 \textit{14} &10\ 17\ 09.95 ,+59\ 32\ 13.8&     19.37&  82768&  44\\
 15 &10\ 17\ 10.25 ,+59\ 37\ 03.8&     20.34&  84579& 101\\
 16 &10\ 17\ 13.36 ,+59\ 35\ 20.6&     19.15&  87734&  83\\
\textit{17} &10\ 17\ 13.69 ,+59\ 37\ 41.9&     20.49&  90966&  94\\
 18 &10\ 17\ 15.97 ,+59\ 31\ 55.8&     20.01&  86089&  99\\
 19 &10\ 17\ 16.35 ,+59\ 37\ 19.0&     20.09&  86157& 113\\
 20 &10\ 17\ 17.61 ,+59\ 35\ 52.8&     19.50&  84662&  74\\
 21 &10\ 17\ 19.36 ,+59\ 34\ 49.8&     19.84&  84626&  64\\
 22 &10\ 17\ 19.53 ,+59\ 34\ 57.1&     19.66&  84213&  63\\
 23 &10\ 17\ 19.60 ,+59\ 34\ 25.7&     19.09&  84922&  87\\
 24 &10\ 17\ 20.53 ,+59\ 37\ 22.0&     18.69&  88760&  87\\
\textit{25} &10\ 17\ 20.74 ,+59\ 35\ 58.8&     17.82&  44148& 119\\
\textit{26} &10\ 17\ 20.82 ,+59\ 31\ 08.7&     17.35&  44735&  56\\
 27 &10\ 17\ 21.99 ,+59\ 34\ 04.5&     19.91&  86645&  74\\
 28 &10\ 17\ 23.77 ,+59\ 36\ 53.1&     19.36&  84216&  59\\
\textit{29} &10\ 17\ 24.30 ,+59\ 30\ 20.0&     20.63&  95482&  90\\
 30 &10\ 17\ 24.45 ,+59\ 37\ 13.6&     20.02&  84715&  83\\
 31 &10\ 17\ 24.99 ,+59\ 30\ 00.4&     20.64&  85806& 100\\
 32 &10\ 17\ 25.29 ,+59\ 31\ 10.1&     19.15&  86995& 102\\
\textit{33} &10\ 17\ 26.11 ,+59\ 37\ 25.4&     19.14& 105530&  53\\
 34 &10\ 17\ 26.47 ,+59\ 37\ 55.8&     19.82&  87113&  60\\
 35 &10\ 17\ 27.50 ,+59\ 35\ 32.9&     19.06&  85702&  63\\
 36 &10\ 17\ 27.74 ,+59\ 35\ 47.2&     20.38&  85061& 108\\

               \noalign{\smallskip}			    
            \hline					    
            \noalign{\smallskip}			    
            \hline					    
         \end{array}
     $$ 
\begin{list}{}{}  
\item[$^{\mathrm{a}}$] IDs in italics indicate non--cluster
galaxies. Galaxy IDs. 53 and 82 (in boldface) indicate the first and
the second dominant galaxies of the cluster (I and II; see text),
respectively.
\end{list}
         \end{table}
\addtocounter{table}{-1}
\begin{table}[!ht]
          \caption[ ]{Continued.}
     $$ 
           \begin{array}{r c c r r}
            \hline
            \noalign{\smallskip}
            \hline
            \noalign{\smallskip}

\mathrm{ID} & \mathrm{\alpha},\mathrm{\delta}\,(\mathrm{J}2000)  & r^\prime & \mathrm{v}\,\,\,\,\,\,\, & \mathrm{\Delta}\mathrm{v}\\
%  &                      &  &  &\,\,\,\,\,\,\,\mathrm{(\,km}&\mathrm{s^{-1}\,)}\,\,\,&  & & \\
  & & &\mathrm{(\,km}&\mathrm{s^{-1}\,)}\\
            \hline
            \noalign{\smallskip}

 37 &10\ 17\ 27.86 ,+59\ 35\ 25.0&     19.95&  86259&  64  \\
 38 &10\ 17\ 28.95 ,+59\ 35\ 24.6&     19.11&  85715&  46  \\
 39 &10\ 17\ 29.35 ,+59\ 30\ 14.2&     19.66&  86304& 115  \\
 40 &10\ 17\ 29.56 ,+59\ 33\ 28.8&     19.96&  88686& 127  \\
 41 &10\ 17\ 29.94 ,+59\ 32\ 31.2&     20.28&  85677& 132  \\
 42 &10\ 17\ 30.47 ,+59\ 34\ 55.9&     19.34&  85747&  64  \\
 \textit{43} &10\ 17\ 31.11 ,+59\ 36\ 59.5&     19.27&  77377&  97  \\
 \textit{44} &10\ 17\ 31.53 ,+59\ 35\ 07.6&     19.57&  54455& 113  \\
 45 &10\ 17\ 31.67 ,+59\ 33\ 54.5&     19.65&  85246& 115        \\
 46 &10\ 17\ 31.96 ,+59\ 31\ 35.9&     18.50&  86766&  77        \\
 47 &10\ 17\ 32.04 ,+59\ 32\ 36.9&     19.76&  87602&  90        \\
 48 &10\ 17\ 32.15 ,+59\ 34\ 24.1&     18.37&  86537&  74        \\
\textit{49} &10\ 17\ 32.73 ,+59\ 30\ 45.3&     19.30&  76879& 109  \\
 50 &10\ 17\ 32.91 ,+59\ 33\ 39.9&     18.33&  84854&  62        \\
 51 &10\ 17\ 33.27 ,+59\ 35\ 12.1&     19.85&  86761&  99        \\
 52 &10\ 17\ 33.55 ,+59\ 34\ 32.7&     18.44&  85846&  35        \\
 \textbf{53} &10\ 17\ 34.34 ,+59\ 33\ 39.5&     17.73&  86351&  49  \\
 54 &10\ 17\ 35.46 ,+59\ 34\ 14.4&     19.18&  86913&  71  \\
 55 &10\ 17\ 35.48 ,+59\ 34\ 58.2&     20.23&  87753& 139  \\
 56 &10\ 17\ 35.96 ,+59\ 34\ 01.6&     18.24&  86057&  73  \\
 57 &10\ 17\ 36.05 ,+59\ 33\ 03.8&     20.02&  86119&  80  \\
 58 &10\ 17\ 36.48 ,+59\ 33\ 16.8&     20.22&  83538&  88  \\
 59 &10\ 17\ 36.65 ,+59\ 34\ 22.9&     19.32&  85483&  98  \\
 60 &10\ 17\ 38.84 ,+59\ 34\ 08.0&     18.94&  86872&  85  \\
 61 &10\ 17\ 39.48 ,+59\ 31\ 38.0&     19.49&  87919&  59  \\
 62 &10\ 17\ 39.48 ,+59\ 32\ 19.2&     20.12&  88109& 113  \\
 \textit{63} &10\ 17\ 39.74 ,+59\ 31\ 23.8&     20.05&  90900&  84  \\
 64 &10\ 17\ 39.90 ,+59\ 34\ 41.9&     20.13&  88277&  99  \\
 65 &10\ 17\ 40.57 ,+59\ 37\ 22.5&     19.95&  85881& 102  \\
 66 &10\ 17\ 40.64 ,+59\ 34\ 50.8&     20.01&  87548&  64  \\
 67 &10\ 17\ 41.09 ,+59\ 35\ 41.0&     19.66&  86020&  66  \\
 68 &10\ 17\ 41.21 ,+59\ 32\ 37.1&     20.29&  85109&  85  \\
 69 &10\ 17\ 42.27 ,+59\ 32\ 53.6&     18.80^{\mathrm{b}}&  85254&  77  \\
 70 &10\ 17\ 42.64 ,+59\ 32\ 11.3&     19.43&  86156&  37  \\
 71 &10\ 17\ 42.72 ,+59\ 36\ 09.7&     19.99&  83344&  63  \\
 72 &10\ 17\ 43.36 ,+59\ 30\ 57.0&     19.75&  85943&  42  \\

                \noalign{\smallskip}			    
            \hline					    
            \noalign{\smallskip}			    
            \hline					    
         \end{array}
     $$ 
\begin{list}{}{}  
\item[$^{\mathrm{b}}$] A visual inspection of the SDSS $r'$ image suggests that the $r'$ magnitude of galaxy ID~69 listed in the SDSS catalog is incorrect, possibly due to the contamination of a very close galaxy. We recompute the $r'$ of ID~69 and present our estimate here.  
\end{list}

         \end{table}

\addtocounter{table}{-1}
\begin{table}[!ht]
          \caption[ ]{Continued.}
     $$ 
           \begin{array}{r c c r r}

            \hline
            \noalign{\smallskip}
            \hline
            \noalign{\smallskip}

\mathrm{ID} & \mathrm{\alpha},\mathrm{\delta}\,(\mathrm{J}2000)  & r^\prime & \mathrm{v}\,\,\,\,\,\,\, & \mathrm{\Delta}\mathrm{v}\\
%  &                      &  &  &\,\,\,\,\,\,\,\mathrm{(\,km}&\mathrm{s^{-1}\,)}\,\,\,&  & & \\
  & & &\mathrm{(\,km}&\mathrm{s^{-1}\,)}\\
            \hline
            \noalign{\smallskip} 

 \textit{73} &10\ 17\ 43.86 ,+59\ 36\ 29.2 &    20.54& 151288&  80    \\	
 74 &10\ 17\ 44.22 ,+59\ 35\ 05.4 &    20.18&  84540& 109   \\	  
 75 &10\ 17\ 44.56 ,+59\ 33\ 58.9 &    19.95&  87489&  66   \\	  
 \textit{76} &10\ 17\ 44.60 ,+59\ 35\ 42.3 &    20.55&  77267&  41   \\	  
 \textit{77} &10\ 17\ 44.76 ,+59\ 33\ 50.3 &    18.06&  42161&  45    \\	  
 78 &10\ 17\ 44.95 ,+59\ 32\ 55.8 &    19.29&  87839&  94    \\	  
 \textit{79} &10\ 17\ 45.00 ,+59\ 35\ 56.7 &    17.64&  4768 &  30   \\	  
 80 &10\ 17\ 45.91 ,+59\ 34\ 18.2 &    19.80&  83828&  52    \\	  
 81 &10\ 17\ 46.00 ,+59\ 33\ 10.2 &    20.17&  86736&  49    \\	  
 \textbf{82} &10\ 17\ 46.03 ,+59\ 32\ 49.7 &    17.84&  86936&  48    \\	  
 83 &10\ 17\ 46.33 ,+59\ 36\ 33.1 &    20.26&  86291& 144    \\	  
 84 &10\ 17\ 46.93 ,+59\ 36\ 34.2 &    19.92&  88762&  72    \\	  
 85 &10\ 17\ 48.14 ,+59\ 32\ 58.6 &    19.71&  86077&  67    \\	  
 86 &10\ 17\ 48.36 ,+59\ 34\ 33.6 &    20.18&  88124&  57    \\	  
 87 &10\ 17\ 50.24 ,+59\ 34\ 34.9 &    19.91&  87569& 102    \\	  
 88 &10\ 17\ 50.34 ,+59\ 35\ 16.3 &    18.09&  86424&  70    \\	  
 89 &10\ 17\ 50.78 ,+59\ 37\ 09.5 &    20.17&  87708& 165    \\	  
\textit{90} &10\ 17\ 50.85 ,+59\ 37\ 00.9 &    20.71& 144784&  30    \\	  
 91 &10\ 17\ 52.45 ,+59\ 31\ 38.7 &    19.62&  84519&  67    \\	  
 92 &10\ 17\ 52.80 ,+59\ 32\ 45.4 &    20.32&  87899&  78    \\	  
 93 &10\ 17\ 53.21 ,+59\ 31\ 54.2 &    19.11&  86432&  49    \\	  
\textit{94} &10\ 17\ 53.81 ,+59\ 31\ 50.5 &    19.96&  99191& 190    \\	  
\textit{95} &10\ 17\ 54.90 ,+59\ 36\ 33.9 &    18.38&  77025&  98     \\	  
 96 &10\ 17\ 55.11 ,+59\ 37\ 47.7 &    19.13&  86790&  77    \\	
 97 &10\ 17\ 56.92 ,+59\ 30\ 18.9 &    19.96&  86714&  91    \\	
 98 &10\ 17\ 59.21 ,+59\ 35\ 34.6 &    20.12&  89084&  78    \\	  
 99 &10\ 17\ 59.98 ,+59\ 36\ 17.3 &    19.52&  88888&  78    \\	
100 &10\ 18\ 00.58 ,+59\ 33\ 36.5 &    19.28&  87585& 118    \\	
\textit{101} &10\ 18\ 00.85 ,+59\ 30\ 36.2 &    20.12& 140381& 105     \\	  
102 &10\ 18\ 03.09 ,+59\ 33\ 28.7 &    19.03&  88458&  85    \\	  
\textit{103} &10\ 18\ 05.46 ,+59\ 37\ 10.0 &    17.95&  34077& 101    \\	  
104 &10\ 18\ 06.71 ,+59\ 36\ 14.5 &    18.16&  88860&  76    \\	
105 &10\ 18\ 07.62 ,+59\ 36\ 08.5 &    18.54&  89358&  76    \\	  
106 &10\ 18\ 10.33 ,+59\ 36\ 47.5 &    19.89&  88541& 144    \\	
\textit{107} &10\ 18\ 11.59 ,+59\ 35\ 41.9 &    17.93&  50182&  67    \\   

              \noalign{\smallskip}			                                                                    
            \hline					    
            \noalign{\smallskip}			    
            \hline					    
         \end{array}\\
     $$ 
         \end{table}

%%
%\end{document}

%% file: tab2.tex
\begin{table}
        \caption[]{Kinematical  properties of galaxy subclumps}
         \label{tabv}
                $$
         \begin{array}{l c r l l}
            \hline
            \noalign{\smallskip}
            \hline
            \noalign{\smallskip}
\mathrm{Clump} & \mathrm{\alpha},\mathrm{\delta}\,(\mathrm{J}2000) &\mathrm{N_g} & \phantom{249}\mathrm{<v>}\phantom{249} & 
\phantom{24}\sigma_{\rm v}^{\mathrm{a}}\phantom{24}\\
&\mathrm{density}\phantom{,}\mathrm{peak}\phantom{,}\mathrm{center} & &\phantom{249}\mathrm{km\ s^{-1}}\phantom{249}&\phantom{2}\mathrm{km\ s^{-1}}\phantom{24}\\
            \hline
            \noalign{\smallskip}
\mathrm{V1^{\mathrm{b}}}     & 10\ 17\ 21.58, +59\ 35\ 10.3 & 20 &84611\pm98 &423_{-66}^{+103}\\
\mathrm{V2^{\mathrm{b}}}     & 10\ 17\ 35.08, +59\ 34\ 17.4 & 37 &86295\pm63 &378_{-34}^{+32}\\
\mathrm{V3^{\mathrm{b}}}     & 10\ 17\ 46.07, +59\ 34\ 20.0 & 24 &88229\pm103&491_{-60}^{+66}\\
\mathrm{KMM1}                & 10\ 17\ 40.87, +59\ 32\ 35.2 & 24 &85464\pm174&831_{-84}^{+127}\\
\mathrm{KMM2}                & 10\ 17\ 34.46, +59\ 34\ 24.7 & 40 &86351\pm153&954_{-71}^{+83}\\
\mathrm{KMM3}                & 10\ 18\ 03.55, +59\ 36\ 15.7 & 17 &88241\pm190&753_{-108}^{+160}\\
\mathrm{DS-NE}               & 10\ 18\ 05.90, +59\ 36\ 14.6 &  6 &88945\pm319&657_{-86}^{+727}\\
              \noalign{\smallskip}
            \hline
            \noalign{\smallskip}
            \hline
         \end{array}
$$
\begin{list}{}{}  
\item[$^{\mathrm{a}}$] We use the biweight estimator and the gapper estimator
by
Beers et al. (1990) for samples with $\mathrm{N_g}\ge$ or $<15$, respectively
(see Girardi et al. \cite{gir93}).
\item[$^{\mathrm{b}}$] The estimate of $\sigma_{\rm V}$ should be
  considered a lower limit in these samples since the membership
  assignment might lead to an artificial truncation of the tails of
  the velocity distributions (e.g., Bird \cite{bir94}).

\
\end{list}
         \end{table}

%% file: tab3.tex
\begin{table*}
        \caption[]{Properties of structures not listed in Table~\ref{tabv}.}
         \label{tabstruct}
                $$
         \begin{array}{l c l l }
            \hline
            \noalign{\smallskip}
            \hline
            \noalign{\smallskip}
\mathrm{Structure} & \mathrm{\alpha},\mathrm{\delta}\,(\mathrm{J}2000) &\phantom{12345678912347}\mathrm{Source} & \phantom{123456789}\mathrm{Additional}\phantom{2}\mathrm{information}\\
& & & \\
            \hline
            \noalign{\smallskip}
\mathrm{WL}\phantom{,}\mathrm{1017.3+5931}& 10\ 17.3, +59\ 31& \mathrm{gravitational}\phantom{,}\mathrm{lensing}\phantom{,}\mathrm{(Dahle}\phantom{,}\mathrm{et}\phantom{,}\mathrm{al.}\phantom{,}\mathrm{\cite{dah03})} &\mathrm{no}\phantom{,}\mathrm{optical}\phantom{,}\mathrm{counterpart}\\
\mathrm{NE}\phantom{2}\mathrm{extension}     & 10\ 18.2, +59\ 37& \mathrm{X}\mathrm{-}\mathrm{ray}\phantom{,}\mathrm{(ROSAT}\phantom{,}\mathrm{archival}\phantom{,}\mathrm{data)}&\mathrm{likely}\phantom{,}\mathrm{connected}\phantom{,}\mathrm{with}\phantom{,}\mathrm{our}\phantom{,}\mathrm{high}\phantom{,}\mathrm{velocity}\phantom{,}\mathrm{clump}\phantom{,}\mathrm{V3}\\
\mathrm{WSW}     & 10\ 16.7, +59\ 32& \mathrm{X}\mathrm{-}\mathrm{ray}\phantom{,}\mathrm{(ROSAT}\phantom{,}\mathrm{archival}\phantom{,}\mathrm{data)}&\mathrm{likely}\phantom{,}\mathrm{infalling}\phantom{,}\mathrm{group}\\
\mathrm{W}                & 10\ 15.9, +59\ 35& \mathrm{2D}\phantom{,}\mathrm{gal.}\phantom{,}\mathrm{distr.}\phantom{,}\mathrm{(SDSS,}\phantom{,}\mathrm{Koester}\phantom{,}\mathrm{et}\phantom{,}\mathrm{al.}\phantom{,}\mathrm{2007)}& \mathrm{MaxBCG}\phantom{,}\mathrm{J}153.93477+59.57870\phantom{,}\mathrm{(z}_{\rm phot}\sim0.281\mathrm{)}\\
\mathrm{far}\phantom{2}\mathrm{NE}                & 10\ 19.2, +59\ 46 & \mathrm{2D}\phantom{,}\mathrm{gal.}\phantom{,}\mathrm{distr.}\phantom{,}\mathrm{(SDSS,}\phantom{,}\mathrm{Koester}\phantom{,}\mathrm{et}\phantom{,}\mathrm{al.}\phantom{,}\mathrm{2007)} &\mathrm{MaxBCG}\phantom{,}\mathrm{J}154.78308+59.76784\phantom{,}\mathrm{(z}_{\rm phot}\sim0.284\mathrm{)}\\
              \noalign{\smallskip}
            \hline
            \noalign{\smallskip}
            \hline
         \end{array}
$$
         \end{table*}

%% file: a959final.bbl
\begin{thebibliography}{}

\bibitem[1989]{abe89} Abell, G. O., Corwin, H. G. Jr., \& Olowin,
R. P. 1989, \apjs, 70, 1

%\bibitem[2000]{arn00} Arnaud, M., Maurogordato, S., Slezak, E., \& Rho, J. 2000, \aap, 355, 461

\bibitem[1994]{ash94} Ashman, K. M., Bird, C. M., \& Zepf, S. E. 1994,
\aj, 108,2348

\bibitem[1994]{bar94} Bardelli, S., Zucca, E., Vettolani, G., et al. 1994, \mnras, 267, 665 

\bibitem[2007a]{bar07a} Barrena, R., Boschin, W., Girardi, M., \& Spolaor, M. 2007a, \aap, 467, 37

\bibitem[2007b]{bar07b} Barrena, R., Boschin, W., Girardi, M., \& Spolaor, M. 2007b, \aap, 469, 861

\bibitem[2005]{bar05} Barrena, R., Ramella, M., Boschin, W., et al. 2005, \aap, 444, 685

\bibitem[1990]{bee90} Beers, T. C., Flynn, K., \& Gebhardt, K. 1990, \aj, 100, 32

\bibitem[1991]{bee91} Beers, T. C., Forman, W., Huchra, J. P., Jones, C., \& Gebhardt, K. 1991, \aj, 102, 1581

\bibitem[1992]{bee92} Beers, T. C., Gebhardt, K., Huchra, J. P., et al. 1992, \apj, 400, 410

%\bibitem[1982]{bee82} Beers, T. C., Geller, M. J., \& Huchra, J. P. 1982, \apj, 257, 23

%\bibitem[1999]{bek99} Bekki, K. 1999, \apj, 510, L15

%\bibitem[1996]{ber96} Bertin, E., \& Arnouts, S. 1996, \aaps, 117, 393

\bibitem[1994]{bir94} Bird, C. M. 1994, \apj, 422, 480

\bibitem[1993]{bir93} Bird, C. M., \& Beers, T. C. 1993, \aj, 105, 1596

%\bibitem[2004]{biv04} Biviano, A., \& Katgert, P. 2004, \aap, 424, 779

\bibitem[2002]{biv02} Biviano, A., Katgert, P., Thomas, T., \& Adami, C. 2002, 
\aap, 387, 8

%\bibitem[1998]{bli98} Bliton, M., Rizza, E., Burns, J. O., Owen, F. N., \& Ledlow, M. J. 1998, \mnras, 301, 609

\bibitem[2002]{boh02} B\"ohringer, H., \& Schuecker, P. 2002, in
``Merging Processes in Galaxy Clusters'', eds. L. Feretti,
I. M. Gioia, \& G. Giovannini (The Netherlands, Kluwer Ac. Pub.):
Observational signatures and statistics of galaxy cluster mergers

%\bibitem[1999]{bor99} Borgani, S., Girardi, M., Carlberg, R. G., Yee, H. K. C., \& Ellingson, E. 1999, \apj, 527, 561

\bibitem[2008]{bos08} Boschin, W., Barrena, R., Girardi, M., \& Spolaor, M. 2008, \aap, 487, 33

\bibitem[2004]{bos04} Boschin, W., Girardi, M., Barrena, R., et al. 2004, \aap, 416, 839

\bibitem[2006]{bos06} Boschin, W., Girardi, M., Spolaor, M., \& Barrena, R. 2006, \aap, 449, 461

%\bibitem[2007]{bra07} Braglia, F., Pierini, D., \& B\"ohringer, H. 2007, \aap, 470, 425

\bibitem[2001]{bul01} Bullock, J. S., Kolatt, T. S., Sigad, Y., et al. 2001, \mnras, 321, 559

\bibitem[2001]{buo01} Buote, D. A. 2001, \apj, 553, 15

\bibitem[2002]{buo02} Buote, D. A. 2002, in ``Merging Processes in
Galaxy Clusters'', eds. L. Feretti, I. M. Gioia, \& G. Giovannini (The
Netherlands, Kluwer Ac. Pub.): Optical Analysis of Cluster Mergers

%\bibitem[1982]{bur82} Burstein, D., \& Heiles, C. 1982, \aj, 87, 1165

\bibitem[1997]{car97} Carlberg, R. G., Yee, H. K. C., \& Ellingson, E. 1997, \apj, 478, 462

%\bibitem[1996]{car96} Carlberg, R. G., Yee, H. K. C., Ellingson, E., et al. 1996, \apj, 462, 32

\bibitem[2006]{cas06} Cassano, R.; Brunetti, G.; \& Setti, G. 2006, \mnras, 369, 1577

\bibitem[1998]{con98} Condon, J. J., Cotton, W. D., Greisen, E. W., et al. 1998, \aj, 115, 1693

\bibitem[1998]{coo98} Cooray, A. R., Grego, L., Holzapfel, W. L., Joy, M., \& Carlstrom, J. E. 1998, \aj, 115, 1388

%\bibitem[2004]{cor04} Cortese, L., Gavazzi, G., Boselli, A., Iglesias--Paramo, J., \& Carrasco, L. 2004, \aap, 425, 429 

%\bibitem[1976]{cou76} Cousins, A. W. J., 1976, Mem. R. Astr. Soc, 81, 25

\bibitem[2002]{dah02} Dahle, H., Kaiser, N., Irgens, R. J., Lilje, P. B., \& 
Maddox, S. J. 2002, \apjs, 139, 313

\bibitem[2003]{dah03} Dahle, H., Pedersen, K., Lilje, P. B., Maddox, S. J., \& Kaiser, N. 2003, \apj, 591, 662

\bibitem[1980]{dan80} Danese, L., De Zotti, C., \& di Tullio, G. 1980, \aap, 
82, 322

%\bibitem[1996]{den96} den Hartog, R., \& Katgert, P. 1996, \mnras, 279, 349

\bibitem[2004]{dol04} Dolag, K., Bartelmann, M., Perrotta, F., et al. 2004, \aap, 416, 853

\bibitem[1988]{dre88} Dressler, A., \& Shectman, S. A. 1988, \aj, 95, 985

%\bibitem[1996]{ebe96} Ebeling, H., Voges, W., B\"ohringer, H., et al. 1996, \mnras, 281, 799

\bibitem[1994]{ell94} Ellingson, E., \& Yee, H. K. C. 1994, \apjs, 92, 33

\bibitem[1996]{fad96} Fadda, D., Girardi, M., Giuricin, G.,
Mardirossian, F., \& Mezzetti, M. 1996, \apj, 473, 670

\bibitem[1987]{fas87} Fasano, G., \& Franceschini, A. 1987, \mnras, 225, 155 

\bibitem[1999]{fer99} Feretti, L. 1999, MPE Report No. 271

\bibitem[2002a]{fer02a} Feretti, L. 2002, The Universe at Low Radio
  Frequencies, Proceedings of IAU Symposium 199, held 30 Nov -- 4 Dec
  1999, Pune, India. Edited by A. Pramesh Rao, G. Swarup, and
  Gopal--Krishna, 2002., p.133

\bibitem[2005]{fer05} Feretti, L. 2005, X--Ray and Radio Connections
(eds. L. O. Sjouwerman and K. K. Dyer). Published electronically by
NRAO, http://www.aoc.nrao.edu/events/xraydio. Held 3--6 February 2004
in Santa Fe, New Mexico, USA

\bibitem[2002b]{fer02b} Feretti, L., Gioia I. M., and Giovannini
G. eds., 2002, Astrophysics and Space Science Library, vol. 272
``Merging Processes in Galaxy Clusters'', Kluwer Academic Publisher,
The Netherlands

%\bibitem[2003]{fer03} Ferrari, C., Maurogordato, S., Cappi, A., \& Benoist C. 2003, \aap, 399, 813

\bibitem[2008]{fer08} Ferrari, C.; Govoni, F.; Schindler, S.; Bykov, A. M.; \& Rephaeli, Y. 2008, \ssr, 134, 93

\bibitem[2000]{flo00} Flores, R. A., Quintana, H., \& Way, M. J. 2000, \apj, 532, 206

%\bibitem[1990]{gio90} Gioia, I. M., Henry, J. P., Maccacaro, T. et al. 1990, \apjl, 356, 35

%\bibitem[1994]{gio94} Gioia, I. M., \& Luppino, G. A. 1994, \apjs, 94, 583 

\bibitem[2002]{gio02} Giovannini, G., \& Feretti, L. 2002, in
``Merging Processes in Galaxy Clusters'', eds. L. Feretti,
I. M. Gioia, \& G. Giovannini (The Netherlands, Kluwer Ac. Pub.):
Diffuse Radio Sources and Cluster Mergers

\bibitem[1999]{gio99} Giovannini, G., Tordi, M., \& Feretti, L. 1999, \na, 4, 141

\bibitem[2008]{gir08} Girardi, M., Barrena, R., Boschin, W., \&
  Ellingson, E. 2008, \aap, 491, 379

\bibitem[2002]{gir02} Girardi, M., \& Biviano, A. 2002, in ``Merging
Processes in Galaxy Clusters'', eds. L. Feretti, I. M. Gioia, \&
G. Giovannini (The Netherlands, Kluwer Ac. Pub.): Optical Analysis of
Cluster Mergers

\bibitem[1993]{gir93} Girardi, M., Biviano, A., Giuricin, G., Mardirossian, F., \& Mezzetti, M. 1993, \apj, 404, 38

\bibitem[2006]{gir06} Girardi, M., Boschin, W., \& Barrena, R. 2006, \aap, 455, 45

%\bibitem[2006]{gir05} Girardi, M., Demarco, R., Rosati, P., \& Borgani, S. 2005, \aap, 442, 29

\bibitem[1996]{gir96} Girardi, M., Fadda, D., Giuricin, G. et al. 1996, 
\apj, 457, 61

\bibitem[1998]{gir98} Girardi, M., Giuricin, G., Mardirossian, F., Mezzetti, M., \& Boschin, W. 1998, \apj, 505, 74

\bibitem[2001]{gir01} Girardi, M., \& Mezzetti, M. 2001, \apj, 548, 79

\bibitem[2002]{got02} Goto, T., Sekiguchi, M., Nichol, R. C., et al. 2002, \aj, 123, 1807

\bibitem[2001a]{gov01a} Govoni, F., Ensslin, T. A., Feretti, L., \& Giovannini, G. 2001a, \aap, 369, 441

\bibitem[2001b]{gov01b} Govoni, F., Feretti, L., Giovannini, G., et al. 2001b, \aap, 376, 803

%\bibitem[2004]{gov04} Govoni, F., Markevitch, M., Vikhlinin, A., et al. 2004, \apj, 605, 695

%\bibitem[1984]{gre84} Gregory, S. A., \& Thompson, L. A. 1984, \apj, 286, 422

%\bibitem[1992]{gul92} Gullixson, C. A. 1992, in ``Astronomical CCD Observing and Reduction techniques'' (ed. S. B. Howell), ASP Conf. Ser., 23, 130

%\bibitem[1995]{hug95} Hughes, J. P., Birkinshaw, M., \& Huchra, J. P. 1995, \apjl, 448, 93

%\bibitem[1953]{joh53} Johnson, H. L., \&  Morgan, W. W. 1953, \apj, 117, 313

\bibitem[2002]{irg02} Irgens, R. J., Lilje, P. B., Dahle, H., \& Maddox, S. J. 2002, 579, 227

\bibitem[2004]{kem04} Kempner, J. C.; Blanton, E. L.; Clarke, T. E.; et al. 2004, in ``The Riddle of Cooling Flows in Galaxies and Clusters of Galaxies'', eds. T. Reiprich, J. Kempner, \& N. Soker (Charlottesville: Univ. Virginia), 335

\bibitem[1992]{ken92} Kennicutt, R. C. 1992, \apjs, 79, 225

\bibitem[2007]{koe07} Koester, B. P., McKay, T. A., Annis, J., et al. 2007, \apj, 660, 239

\bibitem[1982]{led82} Ledermann, W. 1982, Handbook of Applicable Mathematics (New York: Wiley), Vol.6

%\bibitem[1994]{lef94} Le F\`evre, O., Hammer, F., Angonin, M. C., Gioia, I. M., \& Luppino, G. A. 1994, \apjl, 422, 5

\bibitem[1960]{lim60} Limber, D. N., \& Mathews, W. G. 1960, \apj, 132, 286

%\bibitem[1999]{lew99} Lewis, A. D., Ellingson, E., Morris, S. L., \& Carlberg, R. G. 1999, \apj, 517, 587

%\bibitem[1998]{lub98} Lubin, L. M., Postman, M., \& Oke, J. B. 1998, \aj, 116, 643

%\bibitem[2007]{mah07} Mahdavi, A., Hoekstra, H., Babul, A., Balam, D. D., \& Capak, P. L. 2007, \apj, 668, 806 [M07]

\bibitem[1992]{mal92} Malumuth, E. M., Kriss, G. A., Dixon, W. Van Dyke, Ferguson, H. C., \& Ritchie, C. 1992, \aj, 104, 495

%\bibitem[2002]{mar02} Markevitch, M., Gonzalez, A. H., David, L., et al. 2002, \apjl, 567, 27

%\bibitem[2005]{mar05} Markevitch, M., Govoni, F., Brunetti, G, \& Jerius, D. 2005, \apj, 627, 733

%\bibitem[2005]{mat05} Matsuda, Y., Yamada, T., Hayshino, T. et al. 2005, \apjl, 634, 125

%\bibitem[1996]{men96} Menci, N., Fusco--Femiano, R. 1996, \apj, 472, 46

\bibitem[1997]{mus97} Mushotzky, R. F., \& Scharf, C. A. 1997, \apj, 482, L13

\bibitem[1986]{nag86} NAG Fortran Workstation Handbook, 1986 (Downers Grove, IL: Numerical Algorithms Group)

\bibitem[1997]{nav97} Navarro, J. F., Frenk, C. S., \& White, S. D. M. 1997, \apj, 490, 493

%\bibitem[1988]{new88} Newberry, M. V., Kirshner, R. P., \& Boroson, T. A., 1988, \apj, 335, 629

%\bibitem[2008]{oka08} Okabe, N. \& Umetsu, K. 2008, \pasj, 60, 345

\bibitem[2004]{ota04} Ota, N., \& Mitsuda, K. 2004, \aap, 428, 757

\bibitem[1999]{owe99} Owen, F., Morrison, G., \& Voges, W. 1999, proceedings of the workshop ``Diffuse Thermal and Relativistic Plasma in Galaxy Clusters'', eds. H. B\"ohringer, L. Feretti, \& P. Schuecker, MPE Report 271, pp. 9-11

\bibitem[1993]{pis93} Pisani, A. 1993, \mnras, 265, 706

\bibitem[1996]{pis96} Pisani, A. 1996, \mnras, 278, 697

%\bibitem[1997]{pog97} Poggianti, B. M. 1997, \aaps, 122, 399

\bibitem[2004]{pop04} Popesso, P., B\"ohringer, H., Brinkmann, J., Voges, W., \& York, D. 2004, \aap, 423, 449

%\bibitem[1992]{pre92} Press, W. H., Teukolsky, S. A., Vetterling, W. T., \& Flannery, B. P. 1992, in Numerical Recipes (Second Edition), (Cambridge University Press)

%\bibitem[2000]{pro00} Proust, D., Cuevas, H., Capelato, H. V. et al. 2000, \aap, 335, 443

\bibitem[2000]{qui00} Quintana, H., Carrasco, E. R., \& Reisenegger, A. 2000, \aj, 120, 511

\bibitem[2001]{ram01} Ramella, M., Boschin, W., Fadda, D., \& Nonino, M. 2001, \aap, 368, 776

\bibitem[1997]{roe97} Roettiger, K., Loken, C., \& Burns, J. O. 1997, \apjs, 
109, 307

\bibitem[2002]{sar02} Sarazin, C. L. 2002, in ``Merging Processes in
Galaxy Clusters'', eds. L. Feretti, I. M. Gioia, \& G. Giovannini (The
Netherlands, Kluwer Ac. Pub.): The Physics of Cluster Mergers

\bibitem[2001]{sch01} Schuecker, P., B\"ohringer, H., Reiprich, T. H.,
\& Feretti, L. 2001, \aap, 378, 408

%\bibitem[1989]{sha89} Shandarin, S. F. \& Zeldovich Ya. B. 1989, Reviews of Modern Physics, 61, 185

%\bibitem[1965]{sha65} Shapiro, S. S., \& Wilk, M. B. 1965, Biometrika, 52, 591

%\bibitem[2005]{spr05} Springel, V, White, S. D., Jenkins, A. et al., \nat, 435, 629

%\bibitem[1999]{ter99} Terlevich, A. I., Kuntschner, H., Bower, R. G., Caldwell, N., \& Sharples, R. M. 1999, \mnras, 310, 445

\bibitem[1986]{the86} The, L. S., \& White, S. D. M. 1986, \aj, 92, 1248

%\bibitem[1982]{tho82} Thompson, L.A. 1982, in IAU Symposium 104,
%Early Evolution of the Universe and the Present Structure,
%eds. G.O. Abell and G. Chincarini (Dordrecht: Reidel)

\bibitem[1979]{ton79} Tonry, J., \& Davis, M. 1979, \apj, 84, 1511

%\bibitem[1996]{yee96} Yee, H. K. C., Ellingson, E., \& Carlberg, R. G. 1996, \apjs, 102, 269

\bibitem[1978]{wai78} Wainer, H., \&  Schacht, S. 1978, Psychometrika, 43, 203

%\bibitem[2006]{wri06} Wright, E. L. 2006, \pasp, 118, 1711

\end{thebibliography}
